\documentclass[prd,aps,showpacs,nofootinbib,eqsecnum,preprint]{revtex4}
\usepackage{graphicx,bm,color}
\usepackage[small]{caption2}
\usepackage{amsmath}
\usepackage{amsfonts}
\usepackage{amssymb}
\usepackage{amsthm}
\usepackage{amscd}
\usepackage{array}
\usepackage[mathscr]{eucal}
\def\arccot{\mathop{\rm arccot}}
\def\arccoth{\mathop{\rm arccoth}}
\def\csch{\mathop{\rm csch}}

\def\erf{\mathop{\rm erf}}
\def\erfi{\mathop{\rm erfi}}

\def\ei{\mathop{\rm Ei}}

\newcommand{\be}{\begin{equation}}
\newcommand{\ee}{\end{equation}}
\newcommand{\bea}{\begin{eqnarray}}
\newcommand{\eea}{\end{eqnarray}}
\newcommand{\beaa}{\begin{eqnarray*}}
\newcommand{\eeaa}{\end{eqnarray*}}

\newcommand{\e}{\mathrm{e}}

\newcommand{\Eqn}[1]{&\hspace{-0.2em}#1\hspace{-0.2em}&}
\begin{document}
\title{Bounce universe from string-inspired Gauss-Bonnet gravity}
\author{Kazuharu Bamba$^{1, 2}$,
Andrey N. Makarenko$^{3, 4}$,
Alexandr N. Myagky$^{5}$
and
Sergei D. Odintsov$^{6, 7}$
}
\affiliation{
$^1$Leading Graduate School Promotion Center,
Ochanomizu University, 2-1-1 Ohtsuka, Bunkyo-ku, Tokyo 112-8610, Japan\\
$^2$Department of Physics, Graduate School of Humanities and Sciences, Ochanomizu University, Tokyo 112-8610, Japan\\
$^3$Tomsk State Pedagogical University, ul. Kievskaya, 60, 634061 Tomsk, Russia \\
$^4$National Research Tomsk State University, Lenin Avenue, 36, 634050 Tomsk, Russia\\
$^5$National Research Tomsk Polytechnic University, Lenin Avenue, 30,
634050, Tomsk, Russia\\
$^6$Consejo Superior de Investigaciones Cient\'{\i}ficas, ICE/CSIC-IEEC,
Campus UAB, Facultat de Ci\`{e}ncies, Torre C5-Parell-2a pl, E-08193
Bellaterra (Barcelona), Spain\\
$^7$Instituci\'{o} Catalana de Recerca i Estudis Avan\c{c}ats
(ICREA), Barcelona, Spain
}

\begin{abstract}
We explore cosmology with a bounce in Gauss-Bonnet gravity where the 
Gauss-Bonnet invariant couples to a dynamical scalar field. 
In particular, the potential and 
and Gauss-Bonnet coupling function of the scalar field are reconstructed 
so that the cosmological bounce can be realized 
in the case that the scale factor has hyperbolic and exponential forms. 
Furthermore, we examine the relation between 
the bounce in the string (Jordan) and Einstein frames 
by using the conformal transformation between these conformal frames. 
It is shown that in general, the property of the bounce point 
in the string frame changes after the frame is moved to 
the Einstein frame. 
Moreover, it is found that at the point in the Einstein frame 
corresponding to the point of the cosmological bounce in the string frame, 
the second derivative of the scale factor has an extreme value. 
In addition, it is demonstrated that 
at the time of the cosmological bounce in the Einstein frame, 
there is the Gauss-Bonnet coupling function of the scalar field, 
although it does not exist in the string frame. 
\end{abstract}

\hspace{13.0cm} OCHA-PP-329

\maketitle

\def\thesection{\Roman{section}}
\def\theequation{\Roman{section}.\arabic{equation}}


\section{Introduction}

Cosmological observations have suggested that the current cosmic expansion is accelerating. If the current universe is homogeneous, isotropic,
and spatially flat, dark energy with negative pressure or 
the modification of gravity at a large distance is necessary to explain 
the observations (there are recent reviews on the dark energy problem and modified gravity, e.g., in Refs.~\cite{Joyce:2014kja, REV-NO-CF-CD, Bamba:2012cp}). 

On the other hand, 
inflation can explain the homogeneity, isotropy, and flatness of the universe, and include the mechanism to generate the primordial density perturbations. Therefore, inflation is the most promising scenario to describe the early universe. 
As an viable alternative scenario to inflation, 
there has been proposed the matter bounce cosmology~\cite{MBC, Brandenberger:2012zb}, where the initial singularity in the beginning of the universe can be avoided. In this scenario, matter dominates the universe at the bounce point, and the density fluctuations compatible with observations are generated (see, for instance, Ref.~\cite{Novello:2008ra} for a review on bounce cosmology). 
In cosmology with a bounce, there have been various discussions~\cite{Erickson:2003zm} on the BKL instability~\cite{Belinsky:1970ew}, the bounce phenomena~\cite{B-P} in the Ekpyrotic scenario~\cite{Khoury:2001wf}, and the density perturbations~\cite{Cai:2013kja}. Moreover, observational implications of the cosmological bounce have been argued in Ref.~\cite{Prof-Piao}. 

In the matter bounce scenario, 
the primordial density perturbations with a nearly scale-invariant 
and adiabatic spectrum of can be generated~\cite{Brandenberger:2012zb}. 
Especially, the perturbations of the quantum vacuum, whose original scale is smaller than that of the Hubble horizon, are produced. 
Its scale becomes larger than the Hubble horizon 
in the epoch of the contraction where matter dominates the universe, and eventually it evolves as the curvature perturbations with the (almost) scale-invariant spectrum. 
Similarly, it is known that in the Ekpyrotic scenario in the framework 
of brane world models, the primordial density perturbations with such a spectrum can also been produced. One of the most important aims in this scenario is to connect cosmology in the early universe to more fundamental theories such as superstring theories and M-theories~\cite{Khoury:2001wf}. 

Cosmology with a bounce has been examined in various 
gravity theories including $F(R)$ gravity~\cite{B-F(R)},
modified Gauss-Bonnet gravity~\cite{Bamba:2014mya}, 
$f(T)$ gravity~\cite{Cai:2011tc}, where $T$ is the torsion scalar
in teleparallelism, 
non-linear massive gravity with its extension~\cite{Cai:2012ag},
and loop quantum gravity~\cite{Olmo:2008nf, Odintsov:2014gea}
(for references on loop quantum cosmology, see, for example, 
Ref.~\cite{Refs-LQC}). 
The comparison of the bounce cosmology with the BICEP2 experimental data~\cite{Ade:2014xna}\footnote{Very recently, the new joint analysis by BICEP2/Keck Array and Planck~\cite{Ade:2015tva} on $B$-mode polarization and Planck 2015 data~\cite{Planck:2015xua, Ade:2015rim, Ade:2015lrj} on various cosmological aspects have been released.} has been executed in Ref.~\cite{Cai:2014xxa}. 
The parameters of the bounce cosmology with the quasi-matter domination 
have been introduced in Ref.~\cite{Elizalde:2014uba}. 
The theories leading to the cosmological bounce may be represented as a kind of a non-minimal Brans-Dicke-like theory~\cite{A-G}, 
in which anti-gravity behaviours could be realized. 

In this paper, we investigate the cosmological bounce 
in scalar Gauss-Bonnet gravity, where 
a dynamical scalar field non-minimally couples to the Ricci scalar
and/or the Gauss-Bonnet invariant. 
It is known that the Gauss-Bonnet term appears in string theories 
through the approach to derive the low-energy effective action. 
Furthermore, we compare the bounce phenomenon in the string (Jordan) frame 
with that in the Einstein frame by making the conformal transformation 
and explore the relations between these conformal frames. 
We note that the cosmological perturbations~\cite{KS-KSS} and a cosmological scenario for the structure formation~\cite{Kawai:1999xn} in a scalar field theory coupling to the Gauss-Bonnet invariant have been examined. 
Moreover, cosmological non-singular solutions in superstring theories have also been analyzed in Ref.~\cite{Antoniadis:1993jc}. 
We use units of $k_\mathrm{B} = c_{\mathrm{l}} = \hbar = 1$,
where $c$ is the speed of light, and denote the
gravitational constant $8 \pi G$ by
${\kappa}^2 \equiv 8\pi/{M_{\mathrm{Pl}}}^2$
with the Planck mass of $M_{\mathrm{Pl}} = G^{-1/2} = 1.2 \times
10^{19}$\,\,GeV.

The organization of the paper is as follows. 
In Sec.~II, we explain a scalar field theory with non-minimal 
coupling to gravity and derive the equations of motions. 
In the Einstein frame, 
we reconstruct scalar Gauss-Bonnet gravity in Sec.~III 
the Hubble parameter and the scalar field around the cosmological bounce 
in Sec.~IV. 
In Sec.~V, the reconstruction of scalar Gauss-Bonnet gravity is performed
in the string frame. 
In Sec.~VI, we make the conformal transformation of bounce solutions from 
the string frame to the Einstein frame, and vice versa. 
We demonstrate that in general, the bounce in the string frame does not correspond to that in the Einstein frame, and vice versa. 
It is also shown that the bounce universe can be transformed to the accelerating universe in several cases. 
Conclusions are described in Sec.~VII.

\section{Model}

We explore a model of a homogeneous scalar field $\phi=\phi(t)$ non-minimally coupling to gravity. 
Our model action is given by~\cite{Cartier}
\begin{equation}
S=\int d^4x\sqrt{-g}\left\{\frac{1}{2\kappa^2}f(\phi,R)
-\frac{1}{2}\omega(\phi)\left(\nabla\phi\right)^2
-V(\phi)
+\xi(\phi)\left[\alpha_1\mathcal{G}+\alpha_2
\left(\nabla\phi\right)^4
\right]\right\}\,. 
\label{eq:2.1}
\end{equation}
Here, $g$ is the determinant of the metric $g_{\mu\nu}$, 
$\left(\nabla\phi\right)^2 \equiv
g^{\mu\nu}\nabla_{\mu}\phi\nabla_{\nu}\phi$, 
where $\nabla_{\mu}$ is the covariant derivative associated with $g_{\mu\nu}$. 
Moreover, $\mathcal{G}$ is the Gauss-Bonnet invariant 
\begin{equation}
\mathcal{G}=R^2-4R_{\mu\nu}R^{\mu\nu}+R_{\mu\nu\rho\sigma}R^{\mu\nu\rho\sigma}\,,
\label{eq:2.2}
\end{equation}
with $R$ the scalar curvature, $R_{\mu\nu}$ the Ricci tensor,
and $R_{\mu\nu\rho\sigma}$ the Riemann tensor. 
In addition, $f(\phi,R)$ is an arbitrary function of $\phi$ and $R$,
$\omega(\phi)$ and $\xi(\phi)$ are arbitrary functions of $\phi$, 
$V(\phi)$ is the potential of $\phi$, and $\alpha_1$ and $\alpha_2$ are constants. In the following, for simplicity, we set $\kappa^2 =1$. 

Here, we mention that in the framework of string theories 
(for a detailed review, see, e.g.,~\cite{Lidsey:1999mc}), 
the most general expression of the last term 
in the brackets $\{\, \}$ of the action in Eq.~({\ref{eq:2.1}}) 
is represented as~\cite{Metsaev:1987zx} 
\begin{equation}
-\frac{1}{2} \bar{\alpha}' \bar{\lambda} \xi(\phi) 
\left[d_1 \mathcal{G} + d_2 G^{\mu\nu}\nabla_\mu \phi \nabla_\nu \phi 
+ d_3 \Box \phi \left(\nabla\phi\right)^2 + d_4 \left(\nabla\phi\right)^4 \right]\,, 
\label{eq:2.2-2}
\end{equation}
where $\bar{\alpha}' \equiv l_\mathrm{string}^2$ with 
$l_\mathrm{string}$ the fundamental length scale of strings 
is an expansion parameter, $\bar{\lambda}$ is an additional parameter, 
$d_i$ $(i= 1, \dots, 4)$ are constants, 
$G_{\mu\nu} \equiv R_{\mu\nu} -\left(1/2\right)g_{\mu\nu} R$ is the 
Einstein tensor, and 
$\Box \equiv g^{\mu\nu} \nabla_\mu \nabla_\nu$ is the covariant d'Almbertian 
for a scalar field $\phi$. 
Non-singular cosmology~\cite{Copeland:1997ug, FMS-BM} and the cosmological perturbations~\cite{Cartier} in a theory including such higher-order correction terms have been investigated. 
By comparing our action in Eq.~({\ref{eq:2.1}}) with 
the expression in Eq.~({\ref{eq:2.2-2}}), we see that 
in our action in Eq.~({\ref{eq:2.1}}), we have taken $d_2 = 0$ and $d_3 = 0$. 
Furthermore, in the action in Eq.~({\ref{eq:3.1}}) as is shown in the 
next section, we further set $\alpha_1 = 1$ and $\alpha_2 = 0$. Namely, 
we take $-\left(1/2\right) \bar{\alpha}' \bar{\lambda} d_1 =1$ 
and $d_4 = 0$ in the full action in Eq.~({\ref{eq:2.2-2}}). 
This means that our action corresponds to a special model of the general 
theory studied in Ref.~\cite{Cartier}. 
The coefficients $d_i$ have to be determined so that that the full action 
can agree with the three-graviton scattering amplitude~\cite{Cartier}. 

The explanations for the reasons why we have neglected several terms 
in our model action are as follows. 
For the action in Eq.~({\ref{eq:2.1}}), this could be regarded 
as a different version of string-inspired actions from 
the general action including the expression in Eq.~({\ref{eq:2.2-2}}). 
In other words, this action is a kind of a toy model built in 
a phenomenological approach. 
On the other hand, for the action in Eq.~({\ref{eq:3.1}}), 
we consider the case that the non-linear terms of derivatives of 
the scalar field $\phi$, i.e., the last three terms in the action 
in Eq.~({\ref{eq:3.1}}), are much smaller than the Gauss-Bonnet term. 
Therefore, we only take the first term of the action in Eq.~({\ref{eq:2.2-2}}) with $-\left(1/2\right) \bar{\alpha}' \bar{\lambda} d_1 =1$ 
and neglect the other last three terms by setting $d_2 = d_3 = d_4 = 0$

We suppose the flat Friedmann-Lema\^{i}tre-Robertson-Walker (FLRW) metric 
\begin{equation}
ds^2=-dt^2+a^2(t)\sum\limits_{i=1}^{3}(dx^i)^2 \,,
\label{eq:2.3}
\end{equation}
where $a(t)$ is the scale factor.

The variation of the action in Eq.~({\ref{eq:2.1}}) with respect to
the metric $g_{\mu\nu}$ yields the following gravitational equations~\cite{Cartier}
\begin{eqnarray}
\hspace{-12mm}
&&
\frac{1}{2}\omega(\phi)\Dot{\phi}^2
+V(\phi)
+\frac{1}{2}Rf^{\prime}_R(\phi,R)
-\frac{1}{2}f(\phi,R)
-3f^{\prime}_R(\phi,R)H^2
+\frac{1}{2}\rho_c=0 \,,
\label{gb1} \\ 
\hspace{-12mm}
&&
\frac{1}{2}\omega(\phi)\Dot{\phi}^2
-V(\phi)
+\frac{1}{2}f(\phi,R)
-f^{\prime}_R(\phi,R)\left(\Dot{H}+3H^2\right)
\nonumber \\
&&
\hspace{60mm}
{}
+2\Dot{f^{\prime}_R}(\phi,R)H
+\Ddot{f^{\prime}_R}(\phi,R)
+\frac{1}{2}p_c=0 \,. 
\label{gb2}
\end{eqnarray}
By varying the action in Eq.~({\ref{eq:2.1}}) over $\phi$, we obtain 
the equation of motion for $\phi$ as
\begin{equation}
\omega(\phi)\Ddot{\phi}
+3\omega(\phi)H\Dot{\phi}
+V^{\prime}(\phi)
+\frac{1}{2}\omega^{\prime}(\phi)\Dot{\phi}^2
-\frac{1}{2}f^{\prime}_{\phi}(\phi,R)
-\frac{1}{2}\delta_c=0 \,,
\label{gb3}
\end{equation}
with 
\begin{eqnarray}
\rho_c \Eqn{\equiv} -48\alpha_1\xi^{\prime}(\phi)\Dot{\phi}H^3+6\alpha_2\xi(\phi)\Dot{\phi}^4 \,,
\label{eq:2.7} \\
p_c \Eqn{\equiv} 16\alpha_1
\left(H^2\Dot{\phi}^2\xi^{\prime\prime}(\phi)
+H^2\Ddot{\phi}\xi^{\prime}(\phi)
+2H(\Dot{H}+H^2)\Dot{\phi}\xi^{\prime}(\phi)\right)
+2\alpha_2\xi(\phi)\Dot{\phi}^4 \,,
\label{eq:2.8} \\
\delta_c \Eqn{\equiv}
48\alpha_1\xi^{\prime}(\phi)H^2(\Dot{H}+H^2)
-2\alpha_2\Dot{\phi}^2
\left(3\xi^{\prime}(\phi)\Dot{\phi}^2
+12\xi(\phi)\Ddot{\phi}
+12H\xi^{\prime}(\phi)\Dot{\phi}\right) \,.
\label{eq:2.9}
\end{eqnarray}
Here, the dot denotes the time derivative of $d/dt$, 
$f^{\prime}_R(\phi,R) \equiv \partial f(\phi,R)/ \partial R$, 
$f^{\prime}_\phi(\phi,R) \equiv \partial f(\phi,R)/ \partial \phi$,
and the prime shows the derivative operator on a function 
with respect to its argument
as $\omega^{\prime}(\phi) \equiv d \omega(\phi)/d \phi$
and $\xi^{\prime}(\phi) \equiv d \xi(\phi)/d \phi$. 
In the FLRW background in Eq.~({\ref{eq:2.3}}),
the Hubble parameter is defined as $H \equiv \dot{a}/a$.
Furthermore, the scalar curvature and the Gauss-Bonnet invariant
read
$R=6\Dot{H}+12H^2$ and $\mathcal{G}=24H^2 \left(H^2+\Dot{H}\right)$,
respectively. 

It is known that the system of Eqs.~(\ref{gb1}) and (\ref{gb2})
is an overdetermined set of equations. 
We see that Eq.~(\ref{gb2}) is a consequence of Eqs.~(\ref{gb1}) and (\ref{gb3}). By combining Eqs.~(\ref{gb1}) and (\ref{gb2}), we find
\begin{equation}
2f^{\prime}_R\Dot{H}=-\omega(\phi)\Dot{\phi}^2
+H\Dot{f^{\prime}_R}(\phi,R)
-\Ddot{f^{\prime}_R}(\phi,R)
-\frac{1}{2}\rho_c-\frac{1}{2}p_c \,,
\label{eq:2.10}
\end{equation}
or
\begin{eqnarray}
&&
8\alpha_1\Dot{\phi}^2H^2\xi^{\prime\prime}(\phi)
+8\alpha_1(\Ddot{\phi}H^2+2\Dot{\phi}H\Dot{H}-\Dot{\phi}H^3)\xi^{\prime}(\phi)
+4\alpha_2\Dot{\phi}^4\xi(\phi) 
\nonumber \\
&&
=-\omega(\phi)\Dot{\phi}^2
-2f^{\prime}_R(\phi,R)\Dot{H}
+\Dot{f^{\prime}_R}(\phi,R)H
-\Ddot{f^{\prime}_R}(\phi,R) \,.
\label{gb4}
\end{eqnarray}
If the scalar field $\phi(t)$ and the scale factor $a(t)$ are given, 
the coupling function $\xi(\phi)$ may be obtained by solving the differential equation (\ref{gb4}). Hence, the potential of the scalar field $V(\phi)$
can be acquired from Eq.~(\ref{gb1}). 

As an example, we investigate 
the special case that $\alpha_1=1$ and $\alpha_2=0$. 
In this case, the differential equation for $\xi(\phi)$ is represented as 
\begin{equation}
\xi^{\prime\prime}(\phi(t))\Dot{\phi}^2(t)H^2(t)
+\xi^{\prime}(\phi(t))\left(\Ddot{\phi}(t)H^2(t)+2\Dot{\phi}(t)H(t)\Dot{H}(t)-\Dot{\phi}(t)H^3(t)\right)=
a(t)\frac{d}{dt}\left(\frac{H^2(t)}{a(t)}\Dot{\xi}(\phi(t))\right)\,. 
\end{equation}
The solution becomes 
\begin{eqnarray}
\xi(\phi) \Eqn{=}
c_2+c_1\int\limits\frac{a(t)}{H^2(t)}dt
\nonumber \\
&&
{}-\frac{1}{8}\int\limits dt\frac{a(t)}{H^2(t)}
\int\limits dt_1\frac{1}{a(t_1)}
\left(\omega(\phi(t_1))\Dot{\phi}^2(t_1)
+2f^{\prime}_R(\phi(t_1),R(t_1))\Dot{H}(t_1)
\right.
\nonumber \\
&&
\left.\left.
-\Dot{f^{\prime}_R}(\phi(t_1),R(t_1))H(t_1)
+\Ddot{f^{\prime}_R}(\phi(t_1),R(t_1))
\right)
\right|_{t=t(\phi)}\,,
\label{gb4_1}
\end{eqnarray} 
where $c_1$ and $c_2$ are constants. 
Moreover, from Eq.~(\ref{gb1}), we find 
\begin{eqnarray}
V(\phi)\Eqn{=}
24c_1\Dot{a}(t)
-\frac{1}{2}\omega(\phi(t))\Dot{\phi}^2(t)
\nonumber \\
&&
{}+\frac{1}{2}f(\phi(t),R(t))
-3f^{\prime}_R(\phi(t),R(t))\left(\Dot{H}(t)+H^2(t)\right)
+3\Dot{f^{\prime}_R}(\phi(t),R(t))H(t)
\nonumber \\
&&
{}-3\Dot{a}(t)
\int\limits dt_1\frac{1}{a(t_1)}
\left(\omega(\phi(t_1))\Dot{\phi}^2(t_1)
+2f^{\prime}_R(\phi(t_1),R(t_1))\Dot{H}(t_1)
\right.
\nonumber \\
&&
\left.\left.-\Dot{f^{\prime}_R}(\phi(t_1),R(t_1))H(t_1)
+\Ddot{f^{\prime}_R}(\phi(t_1),R(t_1))\right)
\right|_{t=t(\phi)}\,.
\label{gb4_2}
\end{eqnarray}

\section{Reconstruction of scalar Gauss-Bonnet gravity in the Einstein frame}

In this section, we study the action in Eq.~({\ref{eq:2.1}}) with 
$f(\phi,R) = R$, $\omega(\phi) = \gamma \equiv \pm 1$, 
$\alpha_1=1$, and $\alpha_2=0$, expressed as 
\begin{equation}
S=\int d^4x\sqrt{-g}\left(\frac{1}{2}R
-\frac{1}{2}\gamma
\left(\nabla\phi\right)^2
-V(\phi)
+\xi(\phi)\mathcal{G}\right) \,.
\label{eq:3.1}
\end{equation}
This is an action for string-inspired Gauss-Bonnet gravity. 
We reconstruct several models of scalar Gauss-Bonnet gravity. 
We assume that the time dependence of the scalar field 
has the following form:
\begin{equation}
\phi(t) = \phi_0 t \,,
\label{eq:3.2}
\end{equation}
where $\phi_0$ is a constant.

\subsection{Hyperbolic model}

First, we examine the case that the scale factor is given by 
\begin{equation}
a(t)=\sigma \e^{\lambda t}+\tau \e^{-\lambda t}\,, 
\label{eq:3.3}
\end{equation}
where $\sigma$, $\lambda \, (>0)$, and $\tau$ are constants.
In this model, the Hubble parameter and its time derivative read 
\begin{equation}
H(t)=\lambda\frac{\sigma \e^{\lambda t}-\tau \e^{-\lambda t}}{\sigma \e^{\lambda t}+\tau \e^{-\lambda t}}\,, \quad
\Dot{H}(t)=\frac{4\e^{2\lambda t}\lambda^2\sigma\tau}{\left(\e^{2\lambda t}\sigma+\tau\right)^2}\,.
\label{eq:3.4}
\end{equation}
Only one cosmological bounce happens at the time 
\begin{equation}
t_\mathrm{b}=\frac{\ln\left(\tau/\sigma\right)}{2\lambda}\,,
\quad \sigma\tau>0 \,, 
\label{eq:3.5}
\end{equation}
when we have 
\begin{equation}
H(t_\mathrm{b})=0 \,,
\quad
\Dot{H}(t_\mathrm{b})=\lambda^2 \, (>0) \,.
\label{eq:3.6}
\end{equation}
It follows from Eqs.~(\ref{gb4_1}) and (\ref{gb4_2}) with
$f(\phi,R)=R$, $\omega(\phi)=\gamma=\pm 1$, $\alpha_1=1$,
and $\alpha_2=0$ that $V(\phi)$ and $\xi(\phi)$ are described as 
\begin{eqnarray}
V(\phi) \Eqn{=}
-\frac{1}{2}\gamma\phi_0^2
+24c_1\lambda\left(\e^{\frac{\lambda\phi}{\phi_0}}\sigma-\e^{-\frac{\lambda\phi}{\phi_0}}\tau\right)
\nonumber \\
&&
{}-\frac{3(\lambda^2+\gamma\phi_0^2)}{\sqrt{\sigma\tau}}
\left(\e^{\frac{\lambda\phi}{\phi_0}}\sigma-\e^{-\frac{\lambda\phi}{\phi_0}}\tau\right)
\arctan\left(\e^{\frac{\lambda\phi}{\phi_0}}\sqrt{\frac{\sigma}{\tau}}
\right)\,,
\label{eq:3.7} \\
\xi(\phi) \Eqn{=}
c_2
+c_1\frac{\e^{\frac{2\lambda\phi}{\phi_0}}\sigma^2-6\sigma\tau+\e^{\frac{-2\lambda\phi}{\phi_0}}\tau^2}
{\lambda^3\left(\e^{\frac{\lambda\phi}{\phi_0}}\sigma-\e^{-\frac{\lambda\phi}{\phi_0}}\tau\right)}
+\frac{\gamma\phi\phi_0}{8\lambda^3}
\nonumber \\
&&
{}-\frac{\lambda^2+\gamma\phi_0^2}{8\lambda^4\sqrt{\sigma\tau}}
\frac{\e^{\frac{2\lambda\phi}{\phi_0}}\sigma^2-6\sigma\tau+\e^{\frac{-2\lambda\phi}{\phi_0}}\tau^2}
{\e^{\frac{\lambda\phi}{\phi_0}}\sigma-\e^{-\frac{\lambda\phi}{\phi_0}}\tau}
\arctan\left(\e^{\frac{\lambda\phi}{\phi_0}}\sqrt{\frac{\sigma}{\tau}}\right)
\nonumber \\
&&
{}-\frac{1}{8\lambda^4}
\left[(2\lambda^2+\gamma\phi_0^2)\ln\left|\sigma-\e^{\frac{-2\lambda\phi}{\phi_0}}\tau\right|
-2(\lambda^2+\gamma\phi_0^2)\ln\left(\sigma+\e^{\frac{-2\lambda\phi}{\phi_0}}\tau\right)\right]\,.
\label{eq:3.8}
\end{eqnarray}
Furthermore, in the limit
\begin{equation}
\phi\rightarrow\phi_0\frac{1}{2\lambda}\ln\left(\frac{\tau}{\sigma}\right)\,,
\label{eq:3.9}
\end{equation}
we see that 
\begin{equation}
V(\phi)\rightarrow -\frac{1}{2}\gamma\phi_0^2 \,,
\quad
\xi(\phi)\rightarrow\infty \,,
\quad
\text{and}
\quad 
\xi(\phi)\mathcal{G}\rightarrow 0 \,.
\label{eq:3.10}
\end{equation}
%

\begin{figure}[t]
\begin{center}
\includegraphics[width=7cm,height=5cm]{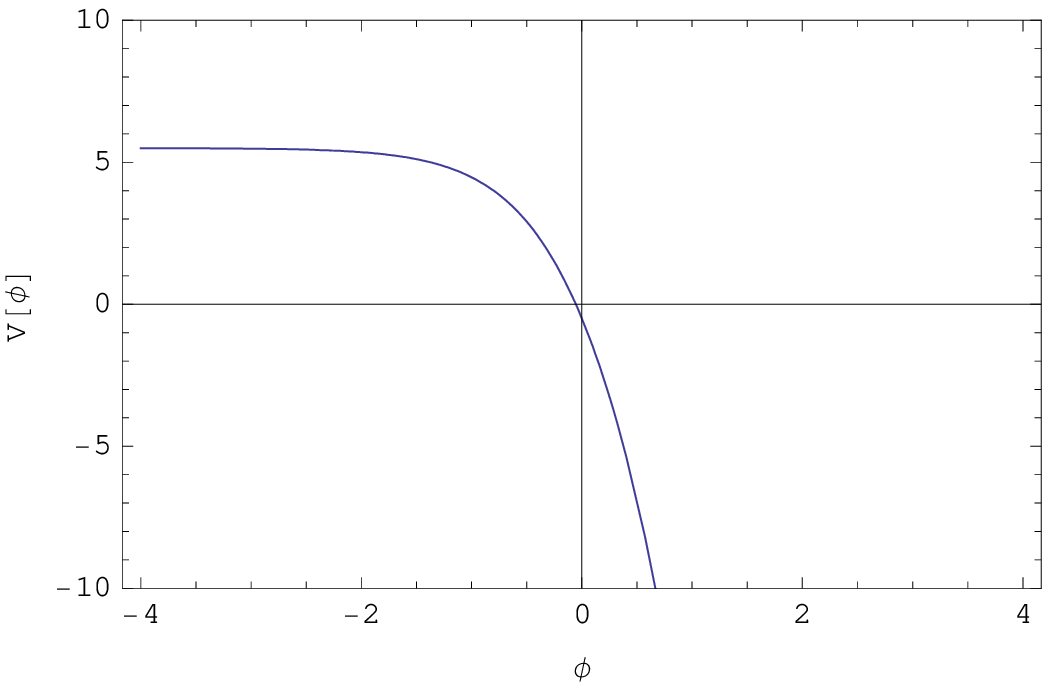}
\hspace{2cm}
\includegraphics[width=7cm,height=5cm]{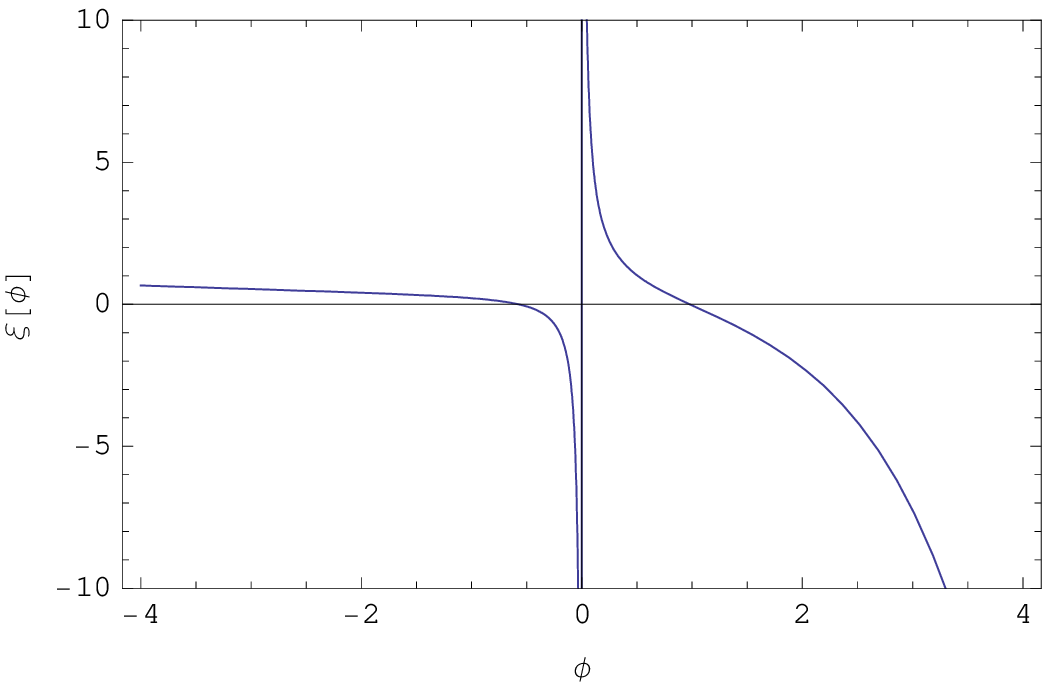}
\end{center}
\caption{
$V(\phi)$ (left panel) and $\xi(\phi)$ (right panel) as functions of $\phi$
for $a(t)=\cosh \lambda t$, 
$c_1=0$, $c_2=0$, $\lambda=1$, $\gamma=1$, and $\phi_0=1$.}
\label{graph_1}
\end{figure}

In Fig.~\ref{graph_1}, we depict the behaviours of $V(\phi)$ and $\xi(\phi)$
as functions of $\phi$ for 
$\sigma=\tau=1/2$, i.e.,
$a(t)=\cosh\lambda t$. For simplicity, we take the following parameter values:
$c_1=0$, $c_2=0$, $\lambda=1$, $\gamma=1$, and $\phi_0=1$.
In this case, we get
\begin{equation}
H(t)=\lambda\tanh\lambda t\,,\quad
\Dot{H}(t)=\frac{\lambda^2}{\cosh^2\lambda t} \,. 
\label{eq:3.11}
\end{equation}
When $t= t_\mathrm{b} = 0$, we find 
\begin{equation}
\quad H(t_\mathrm{b} = 0)=0\,,
\quad
\Dot{H}(t_b = 0)=\lambda^2>0\,.
\label{eq:3.12}
\end{equation}
In addition, if $\phi_0=\lambda$ and therefore $\phi=\lambda t$, 
$V(\phi)$ and $\xi(\phi)$ can be written as
\begin{eqnarray}
V(\phi) \Eqn{=}
-\frac{1}{2}\gamma\lambda^2
+24c_1\lambda\sinh\phi
-6(1+\gamma)\lambda^2\sinh\phi\arctan\e^\phi \,,
\label{eq:3.13} \\
\xi(\phi) \Eqn{=}
c_2 - c_1\frac{1}{\lambda^3}\left(\csch\phi-\sinh\phi\right)
+\frac{1+\gamma}{4\lambda^2}\left(\csch\phi-\sinh\phi\right)
\arctan\e^\phi
\nonumber \\
&&
{}+\frac{1+\gamma}{4\lambda^2}\ln\left(\cosh\phi\right)
-\frac{2+\gamma}{8\lambda^2}\ln\left|\sinh\phi\right|\,.
\label{eq:3.14}
\end{eqnarray}
%

\subsection{Exponential model}

Next, we study the case that the scale factor has an exponential form as
\begin{equation}
a(t)=\exp \left(\alpha t^2\right)\,, 
\label{eq:3.15}
\end{equation}
where $\alpha \, (>0)$ is a positive constant. 
In this case, we have
\begin{equation}
H(t)=2\alpha t \,,
\quad
\Dot{H}(t)=2\alpha \,.
\label{eq:3.16}
\end{equation}
There occurs only one cosmological bounce at $t = t_\mathrm{b} = 0$.
At this time, we obtain 
\begin{equation}
H(t_\mathrm{b} = 0)=0\,,
\quad \Dot{H}(t_\mathrm{b} = 0)=2\alpha \, (>0) \,. 
\label{eq:3.17}
\end{equation}
It follows from Eqs.~(\ref{gb4_1}) and (\ref{gb4_2}) that
\begin{eqnarray}
\hspace{-10mm}
&&
V(\phi) =
\frac{12\alpha^2\phi^2}{\phi_0^2}
+c_1\frac{48\alpha}{\phi_0}\phi \exp\left(\frac{\alpha\phi^2}{\phi_0^2}\right)
-\frac{1}{2}\gamma\phi_0^2
\nonumber \\
\hspace{-10mm}
&&
{}-\frac{3\sqrt{\pi}\sqrt{\alpha}}{\phi_0}\left(4\alpha+\gamma\phi_0^2\right)
\phi \exp\left(\frac{\alpha\phi^2}{\phi_0^2}\right)\erf\left(\frac{\sqrt{\alpha}}{\phi_0}\phi\right)\,,
\label{eq:3.18} \\
\hspace{-10mm}
&&
\xi(\phi) =
c_2 + c_1\frac{\phi_0}{4\alpha^2}
\left(\frac{1}{\phi} \exp\left(\frac{\alpha\phi^2}{\phi_0^2}\right)
-\frac{\sqrt{\pi}\sqrt{\alpha}}{\phi_0}\erfi\left(\frac{\sqrt{\alpha}}{\phi_0}\phi\right)\right)
\nonumber \\
\hspace{-10mm}
&&
{}-\frac{4\alpha+\gamma\phi_0^2}{192\alpha^2\phi_0^2}
\left[2\alpha\phi^2{}_2F_2\left(\{1,1\},\{2,\frac{5}{2}\}; \frac{\alpha\phi^2}{\phi_0^2}\right)
+3\phi_0^2\left(-2+\gamma_\mathrm{E}+\ln\left|\frac{4\alpha\phi^2}{\phi_0^2}\right|\right)\right]\,,
\label{eq:3.19}
\end{eqnarray}
where ${}_2F_2(\sigma_1, \sigma_2, \sigma_3 ;\chi)$ 
with $\sigma_i$ $(i=1, 2, 3)$ constants and $\chi$ a variable 
is a hyper geometric function, 
$\erf(\chi)$ and $\erfi(\chi)$ are the Gauss's error function, 
and $\gamma_\mathrm{E}$ is the Euler's constant.

In the limit that 
\begin{equation}
\phi\rightarrow 0 \,,
\label{eq:3.20}
\end{equation}
we find 
\begin{equation}
V(\phi)\rightarrow -\frac{1}{2}\gamma\phi_0^2 \,,
\quad
\xi(\phi)\rightarrow +\infty \,,
\quad
\text{and}
\quad 
\xi(\phi) \mathcal{G} \rightarrow 0 \,.
\label{eq:3.21}
\end{equation}

\begin{figure}[t]
\begin{center}
\includegraphics[width=7cm,height=5cm]{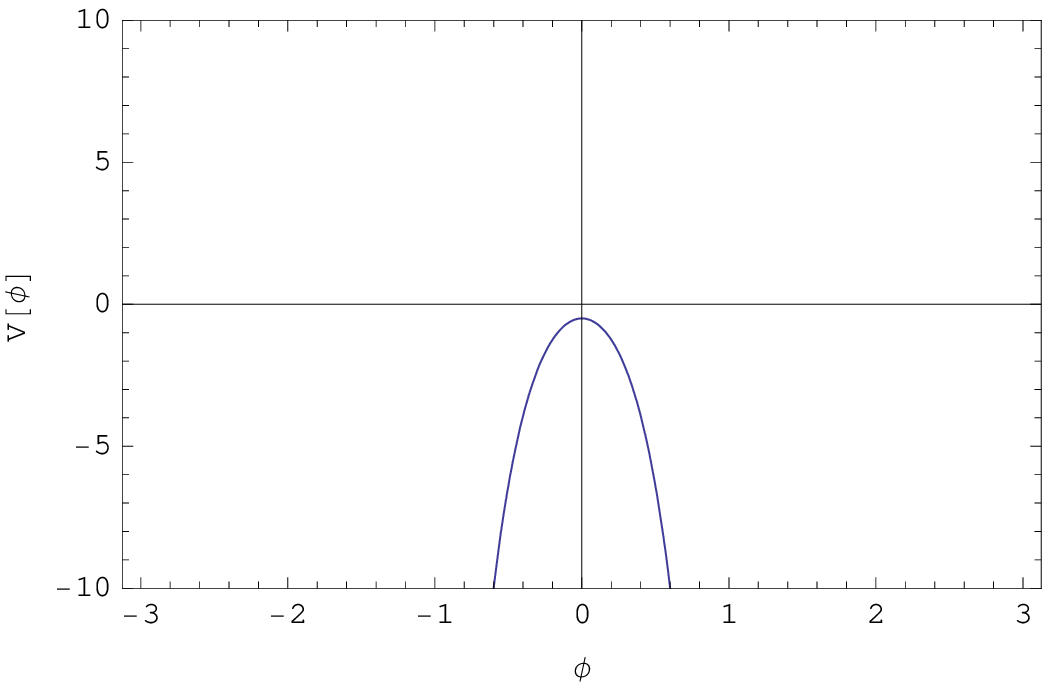}
\hspace{2cm}
\includegraphics[width=7cm,height=5cm]{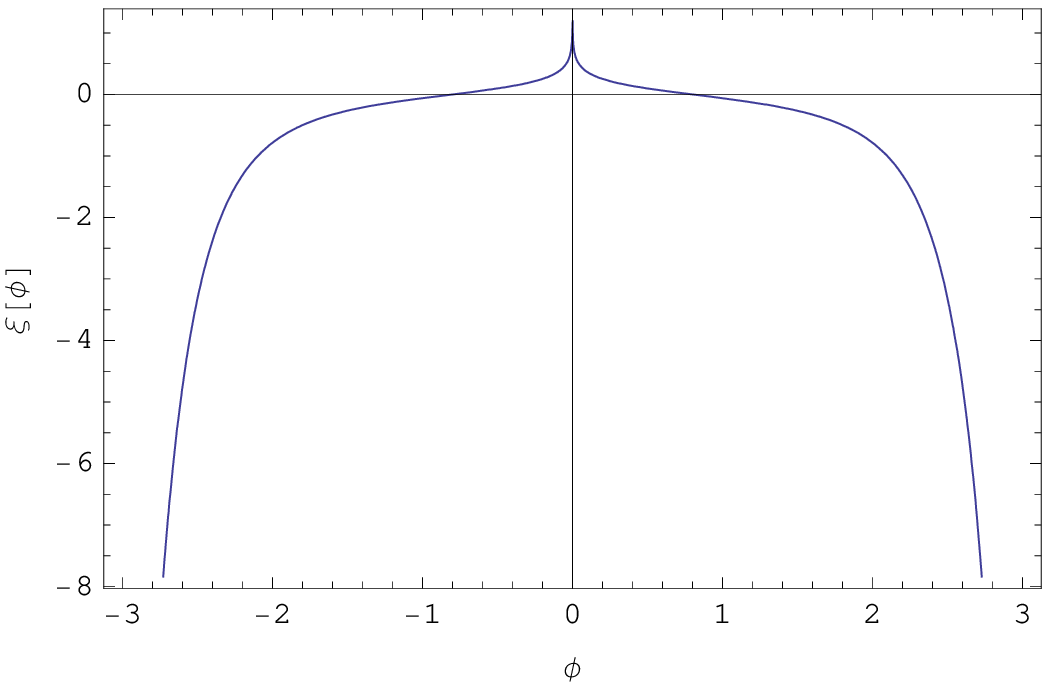}
\end{center}
\caption{$V(\phi)$ (left panel) and $\xi(\phi)$ (right panel)
as functions of $\phi$ for $a(t)=\exp\left(\alpha t^2\right)$
 with $\alpha=1$, $c_1=0$, $c_2=0$, and $\phi_0=1$.}
\label{graph_2}
\end{figure}

In Fig.~\ref{graph_2},
we plot the behaviours of $V(\phi)$ and $\xi(\phi)$ 
as functions of $\phi$ for $a(t)=\exp\left(\alpha t^2\right)$ 
in (\ref{eq:3.15}) with $\alpha=1$, 
$c_1=0$, $c_2=0$, $\gamma=1$, and $\phi_0=1$. 
We remark that for the model with
$\phi_0=\pm\sqrt{-4\gamma\alpha}$ and $c_1=0$, 
$V(\phi)$ and $\xi(\phi)$ are given by
\begin{equation}
V(\phi)=2\alpha-3\gamma\alpha\phi^2 \,,
\quad
\xi(\phi)=c_2 \,.
\label{eq:3.22}
\end{equation}
Hence, if $\alpha>0$, it is necessary that $\gamma=-1$. 

In the end of this section, we explicitly state the physical motivation of our procedure to reconstruct the potential $V(\phi)$ of the scalar field $\phi$ and its coupling function $\xi(\phi)$ to the Gauss-Bonnet invariant for specific forms of the evolutions of $\phi$ and the scale factor $a(t)$ by using the gravitational field equations and the equation of motion for $\phi$. 
Basically, in string theories, 
the functions of $V(\phi)$ and $\xi(\phi)$ are known only 
in some approximations with keeping the leading terms. 
Therefore, it is considered to be very interesting and significant subject to study which forms of these functions are predicted by non-singular cosmology such as cosmology with a bounce to avoid the initial singularity in the early universe. 

In addition, we explain the justification of the reconstructed potentials 
$V(\phi)$ of the scalar field $\phi$ in Eqs.~(\ref{eq:3.7}) and (\ref{eq:3.18}) and its coupling function $\xi(\phi)$ to the Gauss-Bonnet invariant. 
First, we examine the physical behaviours of the resultant potential 
$V(\phi)$. In the reconstructed potentials $V(\phi)$ in Eqs.~(\ref{eq:3.7}) for the hyperbolic form of the scale factor $a(t)=\cosh\lambda t$ in Eq.~(\ref{eq:3.3}) with $\sigma = \tau = 1/2$ and $\phi = \phi_0 t$ in Eq.~(\ref{eq:3.2}), 
on the left panel in Fig.~\ref{graph_1}, the behaviour of $V(\phi)$ for 
$c_1=0$, $c_2=0$, $\lambda=1$, $\gamma=1$, and $\phi_0=1$ is drawn. It follows from this graph that the time of the cosmological bounce is $t_\mathrm{b} = 0$, and hence around the bounce point, the value of the potential $V(\phi)$ changes from positive to negative. 
While, in the reconstructed potentials $V(\phi)$ in Eq.~(\ref{eq:3.18}) for the exponential form of the scale factor $\exp \left(\alpha t^2\right)$ in Eq.~(\ref{eq:3.15}) and $\phi = \phi_0 t$ in Eq.~(\ref{eq:3.2}), 
on the left panel in Fig.~\ref{graph_2}, the behaviour of $V(\phi)$ for 
$\alpha=1$, $c_1=0$, $c_2=0$, and $\phi_0=1$ is shown. {}From this plot, we can see that the time of the cosmological bounce is $t_\mathrm{b} = 0$, and therefore around the bounce point, the absolute value of the potential $V(\phi)$ 
becomes a minimum. 

When we consider inflation in the early universe in the theory whose action 
is given by Eq.~(\ref{eq:3.1}) and regard $\phi$ as the inflaton field, 
it is possible to present a phenomenological justification for the potential forms $V(\phi)$ of $\phi$ in Eqs.~(\ref{eq:3.7}) and (\ref{eq:3.18}). 
For the potential shown on the left panel in Fig.~\ref{graph_1}, 
if the initial value of $\phi$ at the inflationary stage is 
$\phi_\mathrm{i} = -O(1)$, the slow-roll inflation could occur because 
the slope of the potential is sufficiently flat. This form resembles 
a kind of the inflaton potential in the so-called new inflation 
models~\cite{Linde:1981mu, Albrecht:1982wi}. 
Moreover, for the potential shown on the left panel in Fig.~\ref{graph_2}, 
if the initial value of $\phi$ at the inflationary stage is 
$\phi_\mathrm{i} \simeq 0$, the potential form is similar to that in the so-called natural (or axion) inflation~\cite{Freese:1990rb}. 
Consequently, it is considered that the reconstructed potential form $V(\phi)$ includes 
the terms which could have physical and cosmological meanings. 

Next, we discuss the justification of the 
coupling function $\xi(\phi)$ of $\phi$ to the Gauss-Bonnet invariant. 
If we consider the effective action with the loop correction~\cite{Antoniadis:1993jc, Callan:1985ia, A-E-N}, in which non-singular cosmological solutions have been derived~\cite{Rizos:1993rt, Easther:1996yd}, 
it is known that the 
coupling function $\xi(\phi)$ includes 
a constant term, a linear term, an exponential term, and 
a logarithmic term in terms of the scalar field $\phi$~\cite{KS-KSS, Kawai:1999xn}. 
Indeed, there exist these terms in the expressions of $\xi(\phi)$ in Eqs.~(\ref{eq:3.8}) and (\ref{eq:3.19}). For $\xi(\phi)$ in Eq.~(\ref{eq:3.8}), 
the first term is a constant term, the third term is a linear term, 
the second and fourth terms are exponential terms, 
and the fifth term is a logarithmic term. 
Moreover, for $\xi(\phi)$ in Eq.~(\ref{eq:3.19}), 
the first term and the first and second terms within the round 
brackets in the last term on the last line 
are constant terms, the third term is a linear term, 
the second term is an exponential term, 
and the last term within the round 
brackets in the last term on the last line is a logarithmic term. 
Thus, it is interpreted that the resultant coupling function $\xi(\phi)$ 
could consist of the terms whose existence is justified based on the 
considerations in terms of the loop-corrected effective action. 

We also describe the behaviour of $\xi(\phi)$ when the scalar field $\phi$ increases. On the right panel in Fig.~\ref{graph_1}, we see that 
in the limit that $\phi \to +\infty$, $\xi(\phi) \to -\infty$ . 
Moreover, on the right panel in Fig.~\ref{graph_2}, we find that 
$\phi \to \pm \infty$, $\xi(\phi) \to -\infty$. 
Thus, when $\phi$ grows, $\xi(\phi)$ diverges. 
We remark that even if $\xi(\phi)$ diverges, 
$V(\phi)$ can take a finite value, and hence its contribution to the action can also be finite. Thus, there does not occur any divergence related to the divergence of $\xi(\phi)$. 
This fact can be understood by the following investigations. In general, the values of the functions $V(\phi)$ and $\xi(\phi)$ are not limited. This fact is illustrated in Figs.~\ref{graph_1} and \ref{graph_2}. (Even for scalar Gauss-Bonnet gravity in the string frame as is written in Sec.~V, this point can be confirmed in Figs.~~\ref{graph_3} and \ref{graph_4}). 
On the other hand, the function $\xi(\phi)$ has a singularity. 
For example, for the function $\xi(\phi)$ in Eq.~(\ref{eq:3.8})
 and the hyperbolic model of the scale factor $a(t)$ in Eq.~(\ref{eq:3.3}), 
in the limit~(\ref{eq:3.9}), the singularity of $\xi(\phi)$ appears. 
In this limit, the function $V(\phi)$ is finite and the product 
$\xi(\phi) \mathcal{G}$ tends to zero as seen in~(\ref{eq:3.10}). 
The similar behaviour can be found for the function $\xi(\phi)$ 
in Eq.~(\ref{eq:3.19}) and the exponential form of the scale factor $\exp \left(\alpha t^2\right)$ in Eq.~(\ref{eq:3.15}). 
In the limit~(\ref{eq:3.20}), the singularity of $\xi(\phi)$ emerges. 
In this limit, the function $V(\phi)$ is finite and the product 
$\xi(\phi) \mathcal{G}$ tends to zero as represented in~(\ref{eq:3.21}). 
As a result, the behaviour of the function $\xi(\phi)$ does not lead to the appearance of a singularity in the action. 

Furthermore, we explore the qualitative behaviour of $V(\phi)$ in the limit of 
very large or very small values of $\phi$ and $\xi(\phi)$. 
For instance, from Figs.~\ref{graph_1} and \ref{graph_2}, we can see that 
in the limit of the very large (small) value of $\phi$, the absolute value of $V(\phi)$ approaches a very large (small) value. 
(The similar behaviour can be seen also for scalar Gauss-Bonnet gravity in the string frame as is described in Sec.~V. This point can be understood in Figs.~~\ref{graph_3} and \ref{graph_4}). 
Moreover, when the absolute value of $\xi(\phi)$ is very small, $V(\phi)$ takes a finite value, whereas if the (positive) value of $\xi(\phi)$ is very large, 
$V(\phi)$ takes a finite value. 
(The similar behaviour can also be obtained for scalar Gauss-Bonnet gravity in the string frame as is shown in Sec.~V. This phenomena can be seen in Figs.~~\ref{graph_3} and \ref{graph_4}).

\section{Reconstruction of the Hubble parameter and scalar field around 
the cosmological bounce in the Einstein frame}

In this section, we explore the forms of the Hubble parameter and scalar field around the cosmological bounce point in the Einstein frame. 
If the cosmological bounce occurs at the time $t=t_\mathrm{b}$,
the following conditions have to be satisfied
\begin{equation}
H(t_\mathrm{b})=0\,,
\quad
\Dot{H}(t_\mathrm{b})>0\,.
\label{gb5}
\end{equation}
We investigate the Cauchy problem for Eq.~(\ref{gb1}) with $\alpha_1=1$ and
$\alpha_2=0$:
\begin{equation}
\Dot{\phi}=24\gamma H^3\xi^{\prime}(\phi)
\mp\sqrt{6\gamma H^2-2\gamma V(\phi)+576H^6(\xi^{\prime}(\phi))^2}\,,
\quad \phi(t_\mathrm{b})=\phi_\mathrm{b} \,.
\label{eq:4.2}
\end{equation}
The Cauchy problem formulated above takes place in the case that 
\begin{equation}
V(\phi)\leq 0 \quad \text{for} \quad
\gamma=+1 \,,
\label{eq:4.3}
\end{equation}
or
\begin{equation}
V(\phi)\geq 0 \quad \text{for} \quad
\gamma=-1 \,.
\label{eq:4.4}
\end{equation}
We solve the Cauchy problem in the form of the Taylor series 
in the powers of the deviation between the time and the cosmological bounce point $(t-t_\mathrm{b})$: 
\begin{equation}
\phi(t)=\phi_\mathrm{b}+\Dot{\phi}_\mathrm{b}(t-t_\mathrm{b})+\frac{1}{2!}\Ddot{\phi}_\mathrm{b}(t-t_\mathrm{b})^2+\frac{1}{3!}\phi_\mathrm{b}^{(3)}(t-t_\mathrm{b})^3+\cdots \,,
\label{eq:4.5}
\end{equation}
where $\phi_\mathrm{b} = \phi (t=t_\mathrm{b})$ is the value of $\phi$ at 
the bounce point $t_\mathrm{b}$, 
and the superscription $(j)$ $(j = 3, 4, 5)$ 
means the number of the time derivatives. 
Here, we have assumed that the functions $V(\phi)$, $\xi(\phi)$, and $H(t)$
belong to the class $C^{\infty}(\mathcal{U})$, where $\mathcal{U}$
is the vicinity of the cosmological bounce point. 

Equation (\ref{gb1}) and conditions in (\ref{gb5}) lead to the following
relations between the scalar field $\phi(t)$ and Hubble parameter $H(t)$
at the bounce time $t=t_\mathrm{b}$:
\begin{eqnarray}
\hspace{-12mm}
\Dot{\phi_\mathrm{b}}^2 \Eqn{=} -2\gamma V(\phi_\mathrm{b}) \,,
\label{eq:4.5} \\
\hspace{-12mm}
\Ddot{\phi_\mathrm{b}} \Eqn{=} -\gamma V^{\prime}(\phi_\mathrm{b}) \,,
\label{eq:4.6} \\
\hspace{-12mm}
\phi_\mathrm{b}^{(3)} \Eqn{=} \frac{1}{\Dot{\phi_\mathrm{b}}}
\left(
6\gamma\Dot{H}_\mathrm{b}^2
+2V(\phi_\mathrm{b})V^{\prime\prime}(\phi_\mathrm{b})
\right) \,,
\label{gb6} \\
\hspace{-12mm}
\phi_\mathrm{b}^{(4)}
\Eqn{=} 18\gamma\frac{\Dot{H}_\mathrm{b}\Ddot{H}_\mathrm{b}}{\Dot{\phi}_\mathrm{b}}-6\gamma\frac{V^{\prime}(\phi_\mathrm{b})\Dot{H}_\mathrm{b}^2}{V(\phi_\mathrm{b})}+144\gamma\xi^{\prime}(\phi_\mathrm{b})\Dot{H}_\mathrm{b}^3
+V^{\prime}(\phi_\mathrm{b})V^{\prime\prime}(\phi_\mathrm{b})+2V(\phi_\mathrm{b})V^{(3)}(\phi_\mathrm{b})\,,
\label{gb7} \\
\hspace{-12mm}
\phi_b^{(5)} \Eqn{=} \frac{2\gamma}{\Dot{\phi_\mathrm{b}}}
\left[
\frac{27\Dot{H}_\mathrm{b}^4}{V(\phi_\mathrm{b})}
-72\gamma\Dot{H}_\mathrm{b}^3V^{\prime}(\phi_\mathrm{b})\xi^{\prime}(\phi_\mathrm{b})
-\frac{9\gamma\Dot{H}_\mathrm{b}^2(V^{\prime}(\phi_\mathrm{b}))^2}{V(\phi_\mathrm{b})}-576\gamma V(\phi_\mathrm{b})\Dot{H}_\mathrm{b}^3\xi^{\prime\prime}(\phi_\mathrm{b})
\right.
\nonumber \\
\hspace{-12mm}
&&
\left.
{}+432\Dot{H}_\mathrm{b}^2\Ddot{H}_\mathrm{b}\Dot{\phi}_\mathrm{b}\xi^{\prime}(\phi_\mathrm{b})-\frac{27\Dot{H}_\mathrm{b}\Ddot{H}_\mathrm{b}V^{\prime}(\phi_\mathrm{b})\Dot{\phi}_\mathrm{b}}{2V(\phi_\mathrm{b})}
+9\Ddot{H}_\mathrm{b}^2+6\gamma\Dot{H}_\mathrm{b}^2V^{\prime\prime}(\phi_\mathrm{b})+12\Dot{H}_\mathrm{b}\dddot{H}_\mathrm{b}
\right.
\nonumber \\
\hspace{-12mm}
&&
\left.
{}-V(\phi_\mathrm{b})(V^{\prime\prime}(\phi_\mathrm{b}))^2
-3V(\phi_\mathrm{b})V^{\prime}(\phi_\mathrm{b})V^{(3)}(\phi_\mathrm{b})
-2V^2(\phi_\mathrm{b})V^{(4)}(\phi_\mathrm{b})
\vphantom{\frac{27\Dot{H}_\mathrm{b}^4}{V(\phi_\mathrm{b})}}
\right] \,,
\label{gb8}
\end{eqnarray}
where $H_\mathrm{b} = H(t = t_\mathrm{b})$ is the value of $H$ at 
the bounce point $t_\mathrm{b}$, 

Similarly, from Eq.~(\ref{gb4}), we get the values of derivatives of $H(t)$
at the bounce time $t=t_\mathrm{b}$:
\begin{eqnarray}
\hspace{-8mm}
\Dot{H}_\mathrm{b} \Eqn{=} V(\phi_\mathrm{b}) \,,
\label{eq:4.11} \\
\hspace{-8mm}
\Ddot{H}_\mathrm{b} \Eqn{=} \pm\sqrt{-2\gamma V(\phi_\mathrm{b})}\left(V^{\prime}(\phi_\mathrm{b})
-8\xi^{\prime}(\phi_\mathrm{b})V^2(\phi_\mathrm{b})\right),
\label{gb9} \\
\hspace{-8mm}
H_\mathrm{b}^{(3)} \Eqn{=} -6V^2(\phi_\mathrm{b})
-\gamma (V^{\prime}(\phi_\mathrm{b}))^2
-2\gamma V(\phi_\mathrm{b})V^{\prime\prime}(\phi_\mathrm{b})
\nonumber \\
\hspace{-8mm}
&&
{}-384\gamma V^4(\phi_\mathrm{b})(\xi^{\prime}(\phi_\mathrm{b}))^2
+72\gamma V^2(\phi_\mathrm{b})V^{\prime}(\phi_\mathrm{b})\xi^{\prime}(\phi_\mathrm{b})+48\gamma V^3(\phi_\mathrm{b})\xi^{\prime\prime}(\phi_\mathrm{b})\,.
\label{gb10}
\end{eqnarray}
Consequently,
if we have the potential $V(\phi)$ of the scalar field $\phi$ and 
the coupling function of $\phi$ to the Gauss-Bonnet invariant $\xi(\phi)$, 
the expansion of the function $H(t)$ around the bounce time $t=t_\mathrm{b}$
can be written as 
\begin{equation}
H(t)=\Dot{H}_b(t-t_\mathrm{b})+\frac{1}{2!}\Ddot{H}_\mathrm{b}(t-t_\mathrm{b})^2+\frac{1}{3!}H_\mathrm{b}^{(3)}(t-t_\mathrm{b})^3+\cdots
\,.
\label{gb11}
\end{equation}
The condition $\Dot{H}_\mathrm{b}>0 \, (<0)$ leads to the following expression $V(\phi_\mathrm{b})>0 \, (<0)$, but the expansion in Eq.~(\ref{gb11}) is available only if $\gamma=-1 \, (+1)$. 
Through the combination of Eqs.~(\ref{gb6})-(\ref{gb10}), we acquire
\begin{eqnarray}
\hspace{-10mm}
\Dot{\phi_\mathrm{b}} \Eqn{=} \mp\sqrt{-2\gamma V(\phi_\mathrm{b})} \,,
\label{eq:4.15} \\
\hspace{-10mm}
\Ddot{\phi_\mathrm{b}} \Eqn{=} -\gamma V^{\prime}(\phi_\mathrm{b}) \,,
\label{eq:4.16} \\
\hspace{-10mm}
\phi_\mathrm{b}^{(3)} \Eqn{=}
\mp\frac{6\gamma V^2(\phi_\mathrm{b})+2V(\phi_\mathrm{b})V^{\prime\prime}(\phi_\mathrm{b})}{\sqrt{-2\gamma V(\phi_\mathrm{b})}} \,,
\label{eq:4.17} \\
\hspace{-10mm}
\phi_\mathrm{b}^{(4)} \Eqn{=}
12\gamma V(\phi_\mathrm{b})V^{\prime}(\phi_\mathrm{b})
+V^{\prime}(\phi_\mathrm{b})V^{\prime\prime}(\phi_\mathrm{b})
+2V(\phi_\mathrm{b})V^{(3)}(\phi_\mathrm{b})\,,
\label{eq:4.18} \\
\hspace{-10mm}
\phi_\mathrm{b}^{(5)} \Eqn{=}
\mp\sqrt{-2\gamma V(\phi_\mathrm{b})}
\left[
45V^2(\phi_\mathrm{b})
-1152\gamma V^4(\phi_\mathrm{b})(\xi^{\prime}(\phi_\mathrm{b}))^2
+12\gamma (V^{\prime}(\phi_\mathrm{b}))^2
\right.
\nonumber \\
\hspace{-10mm}
&&
\left.
{}+(V^{\prime\prime}(\phi_\mathrm{b}))^2
+3V^{\prime}(\phi_\mathrm{b})V^{(3)}(\phi_\mathrm{b})
+18\gamma V(\phi_\mathrm{b})V^{\prime\prime}(\phi_\mathrm{b})
+2V(\phi_\mathrm{b})V^{(4)}(\phi_\mathrm{b})
\right]\,.
\label{eq:4.19}
\end{eqnarray}
Therefore, the interaction of a scalar field with the Gauss-Bonnet invariant appears in the expansion near the point $t=t_\mathrm{b}$ only from the fifth order.
If the potential $V(\phi)$ and function $\xi(\phi)$ are represented as
\begin{eqnarray}
V(\phi) \Eqn{=} V_0 \exp \left(-\frac{2\phi}{\phi_0} \right)\,,
\label{eq:4.20} \\
\xi(\phi) \Eqn{=} \xi_0 \exp \left(\frac{2\phi}{\phi_0} \right)\,,
\label{eq:4.21}
\end{eqnarray}
with $V_0$, $\xi_0$, and $\phi_0$ constants, 
a scalar field can be expanded near the point of the cosmological
bounce as 
\begin{eqnarray}
\phi(t) \Eqn{=} \phi_\mathrm{b}\mp\sqrt{-2\gamma V_0} \exp \left(-\frac{\phi_\mathrm{b}}{\phi_0}\right)(t-t_\mathrm{b})
+\frac{\gamma V_0}{\phi_0} \exp \left(-2\frac{\phi_\mathrm{b}}{\phi_0}\right)(t-t_\mathrm{b})^2
\nonumber \\
&&
{}\pm\frac{V_0^2(4+3\gamma\phi_0^2)}{3\phi_0^2\sqrt{-2\gamma V_0}} \exp \left(-3\frac{\phi_\mathrm{b}}{\phi_0}\right)(t-t_\mathrm{b})^3
\nonumber \\
&&
{}-\frac{V_0^2}{\phi_0^3}(1+\gamma\phi_0^2) \exp \left(-4\frac{\phi_\mathrm{b}}{\phi_0}\right)(t-t_\mathrm{b})^4 
+\cdots \,.
\label{eq:4.22}
\end{eqnarray}
The expansion of the function $H(t)$ around the bouncing time $t=t_\mathrm{b}$
becomes
\begin{eqnarray}
H(t) \Eqn{=} V_0 \exp\left(-2\frac{\phi_\mathrm{b}}{\phi_0}\right)(t-t_\mathrm{b})\pm\sqrt{-2\gamma V_0}\frac{V_0}{\phi_0}(1+8\xi_0V_0)\exp\left(-3\frac{\phi_\mathrm{b}}{\phi_0}\right)(t-t_\mathrm{b})^2
\nonumber \\
&&
{}-\frac{\gamma V_0^2}{\phi_0^2}
\left[2+16\xi_0V_0(1+16\xi_0V_0)+\gamma\phi_0^2\right] \exp\left(-4\frac{\phi_\mathrm{b}}{\phi_0}\right)(t-t_\mathrm{b})^3
+\cdots \,.
\label{eq:4.23}
\end{eqnarray}
When the expressions of $\phi(t)$ and $H(t)$ are known, it is possible to reconstruct the functions $\xi(\phi)$ and $V(\phi)$ around $t=t_\mathrm{b}$. 
However, only if the form of the term $\xi(\phi)(\nabla\phi)^2$ in the action is taken into account, the function $\xi(\phi)$ can completely be reconstructed. 

Equation (\ref{gb4}) is consistently differentiated with respect to the variable $t$. Accordingly, we find the coefficients of expansion of $\xi(\phi)$ in the Taylor series around the cosmological bounce at $t=t_\mathrm{b}$. 
The Taylor expansion of $\xi(\phi)$ is given by 
\begin{equation}
\xi(\phi)=
\xi(\phi_\mathrm{b})
+\xi^{\prime}(\phi_\mathrm{b})(\phi-\phi_\mathrm{b})
+\frac{1}{2}\xi^{\prime\prime}(\phi_\mathrm{b})(\phi-\phi_\mathrm{b})^2+\cdots
\,.
\label{eq:4.24}
\end{equation}
Here, for $\alpha_1\neq 0$ and $\alpha_2\neq 0$, we have
\begin{equation}
\xi(\phi_\mathrm{b})=
-\frac{2\Dot{H}_\mathrm{b}+\gamma\Dot{\phi}_\mathrm{b}^2}{4\alpha_2\Dot{\phi}_\mathrm{b}^4}\,,
\quad
\xi^{\prime}(\phi_\mathrm{b})=
\frac{(4\Dot{H}_\mathrm{b}+\gamma\Dot{\phi}_\mathrm{b}^2)\Ddot{\phi}_\mathrm{b}-\Dot{\phi}_\mathrm{b}\Ddot{H}_\mathrm{b}}
{8\alpha_1\Dot{H}_\mathrm{b}^2\Dot{\phi}_\mathrm{b}^2+2\alpha_2\Dot{\phi}_\mathrm{b}^6},
\quad
\cdots \,,
\label{eq:4.25}
\end{equation}
whereas 
for $\alpha_1=1$ and $\alpha_2=0$, we obtain 
\begin{equation}
\xi(\phi_\mathrm{b})=c_1 \,,
\quad
\xi^{\prime}(\phi_\mathrm{b})=
-\frac{\Ddot{H}_\mathrm{b}+\gamma\Dot{\phi}_\mathrm{b}\Ddot{\phi}_\mathrm{b}}{8\Dot{H}_\mathrm{b}^2\Dot{\phi}_\mathrm{b}}\,,
\quad
\cdots \,,
\label{eq:4.26}
\end{equation}
where $c_1$ is a constant.

{}From Eq.~(\ref{gb1}), we see that 
the function $V(\phi)$ can be expanded
in the Taylor series around the cosmological bounce time. 
In the hyperbolic model of 
$a(t)=\cosh(\lambda t)$ with $\lambda>0$,
for $\alpha_1\neq 0$ and $\alpha_2\neq 0$,  
if the scalar field is
represented as $\phi(t)=\phi_0 t$ in Eq.~(\ref{eq:3.2}), 
we get 
\begin{eqnarray}
\xi(\phi) \Eqn{=}
-\frac{2\lambda^2+\gamma\phi_0^2}{4\alpha_2\phi_0^4}
+\frac{\lambda^4}{2\phi_0^2(12\alpha_1\lambda^4+\alpha_2\phi_0^4)}\phi^2
+\cdots \,,
\label{eq:4.27} \\
V(\phi) \Eqn{=}
\frac{1}{4}(6\lambda^2+\gamma\phi_0^2)
+\frac{3}{2}\frac{\lambda^4}{\phi_0^2}\left(1+\frac{12\alpha_1\lambda^4}{12\alpha_1\lambda^4+\alpha_2\phi_0^4}\right)\phi^2
+\cdots \,.
\label{eq:4.28}
\end{eqnarray}
In addition, for $\alpha_1=1$ and $\alpha_2=0$, we have
\begin{eqnarray}
\xi(\phi) \Eqn{=} c_1+\frac{1}{24\phi_0^2}\phi^2+\frac{7\lambda^2}{720\phi_0^4}\phi^4 +\cdots \,,
\label{eq:4.29} \\
V(\phi) \Eqn{=} -\frac{1}{2}\gamma\phi_0^2
+\frac{3\lambda^4}{\phi_0^2}\phi^2
+\cdots \,.
\label{eq:4.30}
\end{eqnarray}
On the other hand, in the exponential model of 
$a(t)=\exp\left(\alpha t^2\right)$ with $\alpha>0$ in Eq.~(\ref{eq:3.15}), 
if $\alpha_1\neq 0$ and $\alpha_2\neq 0$, we obtain 
\begin{eqnarray}
\xi(\phi) \Eqn{=}
-\frac{4\alpha+\gamma\phi_0^2}{4\phi_0^4\alpha_2}\,,
\label{eq:4.31} \\
V(\phi) \Eqn{=}
3\alpha+\frac{1}{4}\gamma\phi_0^2+\frac{12\alpha^2}{\phi_0^2}\phi^2 \,.
\label{eq:4.32}
\end{eqnarray}
Furthermore, for $\alpha_1=1$ and $\alpha_2=0$, we acquire
\begin{eqnarray}
\xi(\phi) \Eqn{=} c_1 \,,
\label{eq:4.33} \\
V(\phi) \Eqn{=}
-\frac{1}{2}\gamma\phi_0^2+\frac{12\alpha^2}{\phi_0^2}\phi^2 \,.
\label{eq:4.34}
\end{eqnarray}
%

\section{Reconstruction of scalar Gauss-Bonnet gravity in the string frame}

In Sec.~III, we have examined scalar Gauss-Bonnet gravity in the Einstein 
frame, while in this section, we study scalar Gauss-Bonnet gravity in the string frame. In the string (Jordan) frame, the action has the form 
\begin{equation}
S=\int d^4x\sqrt{-g}
\left\{\e^{-\phi}\left[\frac{1}{2}R
+\frac{1}{2}\left(\nabla\phi\right)^2
-V(\phi)\right]
+\xi(\phi)\mathcal{G}\right\} \,.
\label{eq:5.1}
\end{equation}
We consider the case that the scalar field is expressed as
$\phi(t)=\phi_0 t$ in Eq.~(\ref{eq:3.2}).
It follows from the action in Eq.~(\ref{eq:5.1}) that
in the FLRW space-time in Eq.~(\ref{eq:2.3}),
the gravitational field equations and the equation of motion for $\phi$
read
\begin{eqnarray}
\hspace{-10mm}
&&
6H^2-6H\Dot{\phi}+\Dot{\phi}^2-2V(\phi)=
-48\e^{\phi}\xi^{\prime}(\phi)\Dot{\phi}H^3 \,,
\label{gb12} \\
\hspace{-10mm}
&&
4\Dot{\phi}H-4\Dot{H}-6H^2-\Dot{\phi}^2+2\Ddot{\phi}+2V(\phi)
=\e^{\phi}\left[16f^{\prime\prime}(\phi)\Dot{\phi}^2H^2
\right.
\nonumber \\
\hspace{-10mm}
&&
\hspace{70mm}
\left.
{}+16f^{\prime}(\phi)\Ddot{\phi}H^2
+32f^{\prime}(\phi)\Dot{\phi}H\left(\Dot{H}+H^2\right)\right]\,,
\label{gb13} \\
\hspace{-10mm}
&&
6\Dot{H}+12H^2+\Dot{\phi}^2-2\Ddot{\phi}-6H\Dot{\phi}
-2V(\phi)+2V^{\prime}(\phi)
=48\e^{\phi}\xi^{\prime}(\phi)H^2\left(\Dot{H}+H^2\right)\,.
\label{gb14}
\end{eqnarray}
For the string frame, from Eqs.~(\ref{gb4_1}) and (\ref{gb4_2}) with
$f(\phi,R)=\e^{-\phi}R$, $\omega(\phi)=-\e^{-\phi}$, $\alpha_1=1$,
and $\alpha_2=0$, we have
\begin{eqnarray}
\xi(\phi) \Eqn{=}
\left.
c_2+c_1\int\limits\frac{a(t)}{H^2(t)}dt
\right.
\nonumber \\
&&
\left.
{}+\frac{1}{8}\int\limits dt\frac{a(t)}{H^2(t)}
\int\limits dt_1\frac{\e^{-\phi(t_1)}}{a(t_1)}\left(\Ddot{\phi}(t_1)-H(t_1)\Dot{\phi}(t_1)-2\Dot{H}(t_1)\right)
\right|_{t=t(\phi)}\,,
\label{eq:5.5} \\ 
V(\phi) \Eqn{=}
24c_1\Dot{a}(t)\e^{\phi(t)}
+3H^2(t)+\frac{1}{2}\Dot{\phi}^2(t)-3H(t)\Dot{\phi}(t)
\nonumber \\
&&
\left.
{}+3\Dot{a}(t)\e^{\phi(t)}
\int\limits dt_1\frac{\e^{-\phi(t_1)}}{a(t_1)}\left(\Ddot{\phi}(t_1)-H(t_1)\Dot{\phi}(t_1)-2\Dot{H}(t_1)\right)
\right|_{t=t(\phi)}\,.
\label{eq:5.6}
\end{eqnarray} 
%

\subsection{Hyperbolic model}

When the scale factor is written as
$a(t)=\sigma \e^{\lambda t}+\tau \e^{-\lambda t}$ with $\lambda>0$
in (\ref{eq:3.3}), 
by using the assumption that $\phi(t)=\lambda t$, 
the potential $V(\phi)$ and interaction function $\xi(\phi)$
are reconstructed as
\begin{eqnarray}
V(\phi) \Eqn{=} 24c_1\lambda\left(\e^{2\phi}\sigma-\tau\right)
+\frac{7}{2}\lambda^2
-\frac{6\lambda^2\tau}{\e^{2\phi}\sigma+\tau}
-3\lambda^2\phi
+\frac{3\lambda^2\sigma}{\tau}\phi \e^{2\phi}
\nonumber \\
&&
{}-\frac{3\lambda^2}{2\tau}\left(\e^{2\phi}\sigma-\tau\right)\ln\left(\e^{2\phi}\sigma+\tau\right)\,,
\label{eq:5.7} \\
\xi(\phi) \Eqn{=} c_2
+c_1\frac{\e^{-\phi}\left(\e^{4\phi}\sigma^2-6\e^{2\phi}\sigma\tau+\tau^2\right)}{\lambda^3\left(\e^{2\phi}\sigma-\tau\right)}
\nonumber \\
&&
{}+\frac{\sigma}{2\lambda^2\sqrt{\sigma\tau}}\left(\arccot\left(\e^{\phi}\sqrt{\frac{\sigma}{\tau}}\right)
+\arccoth\left(\e^{\left|\phi\right|}\sqrt{\frac{\sigma}{\tau}}\right)\right)
+\frac{\sigma}{16\lambda^2\tau}\e^{\phi}\left(2\phi-\ln\left(\e^{2\phi}\sigma+\tau\right)\right)
\nonumber \\
&&
{}-\frac{\sigma}{4\lambda^2}\frac{\e^{\phi}\left(2+2\phi-\ln\left(\e^{2\phi}\sigma+\tau\right)\right)}{\e^{2\phi}\sigma-\tau}
+\frac{1}{16\lambda^2}\e^{-\phi}\left(-8-2\phi+\ln\left(\e^{2\phi}\sigma+\tau\right)\right)\,.
\label{eq:5.8}
\end{eqnarray}
Moreover, we see that
\begin{equation}
V(\phi)\rightarrow \frac{1}{2}\lambda^2 \,, \quad
\xi(\phi)\rightarrow\infty, \quad
\xi(\phi)\mathcal{G}\rightarrow 0 \
\quad \text{for} \quad
\phi\rightarrow\frac{1}{2}\ln\left(\frac{\tau}{\sigma}\right)\,.
\label{eq:5.9}
\end{equation}
If we set $\sigma=\tau=1/2$, we obtain
\begin{eqnarray}
V(\phi) \Eqn{=}
24c_1\lambda \e^{\phi}\sinh\phi
+\frac{1}{2}\lambda^2
+3\lambda^2\tanh\phi
\nonumber \\
&&
{}
+6\lambda^2\phi \e^{\phi}\sinh\phi
-3\lambda^2 \e^{\phi}\sinh\phi\ln\left(\e^{\phi}\cosh\phi\right)\,,
\label{eq:5.11} \\
\xi(\phi) \Eqn{=} c_2
+c_1\frac{1}{2\lambda^3}\left(-3+\cosh(2\phi)\right)\csch\phi 
+\frac{1}{16\lambda^2}\e^{\phi}(-1+\coth\phi)
\left[-8-6\phi 
\right.
\nonumber \\
&&
\left.
{}-3\ln2
+\cosh(2\phi)(4+2\phi+\ln2)
-(-3+\cosh(2\phi))\ln\left(1+\e^{2\phi}\right)
\right.
\nonumber \\
&&
\left.
{}+8\left(\arccot \e^{\phi}+\arccoth \e^{|\phi|}\right)\sinh\phi
-4\sinh(2\phi)\right]\,.
\label{eq:5.12}
\end{eqnarray}
%

\begin{figure}[t]
\begin{center}
\includegraphics[width=7cm,height=5cm]{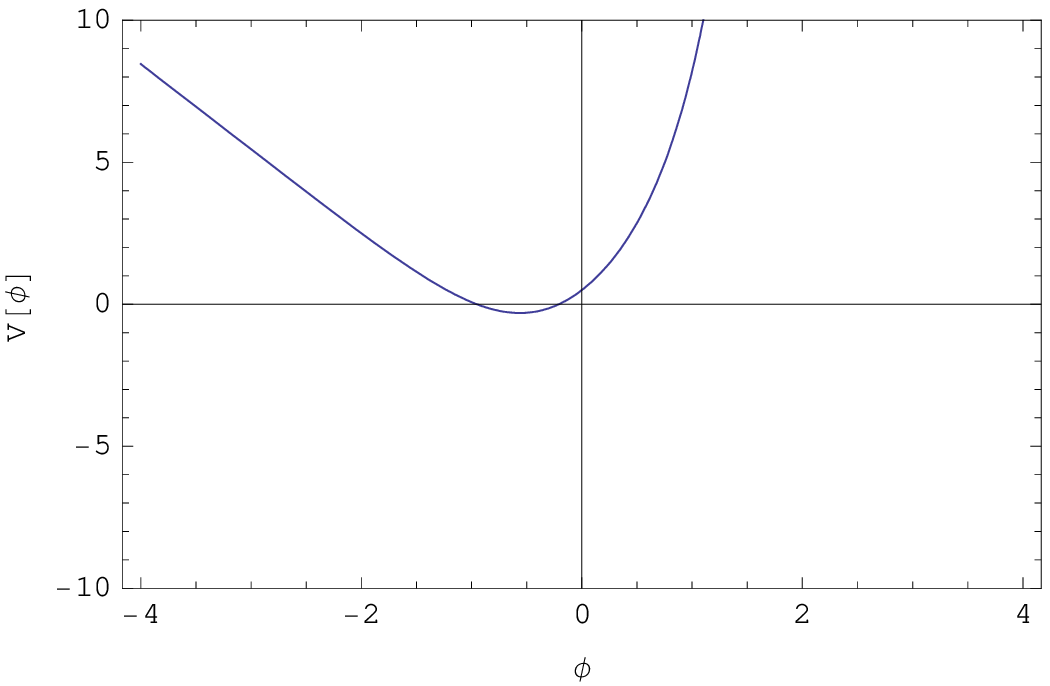}
\hspace{2cm}
\includegraphics[width=7cm,height=5cm]{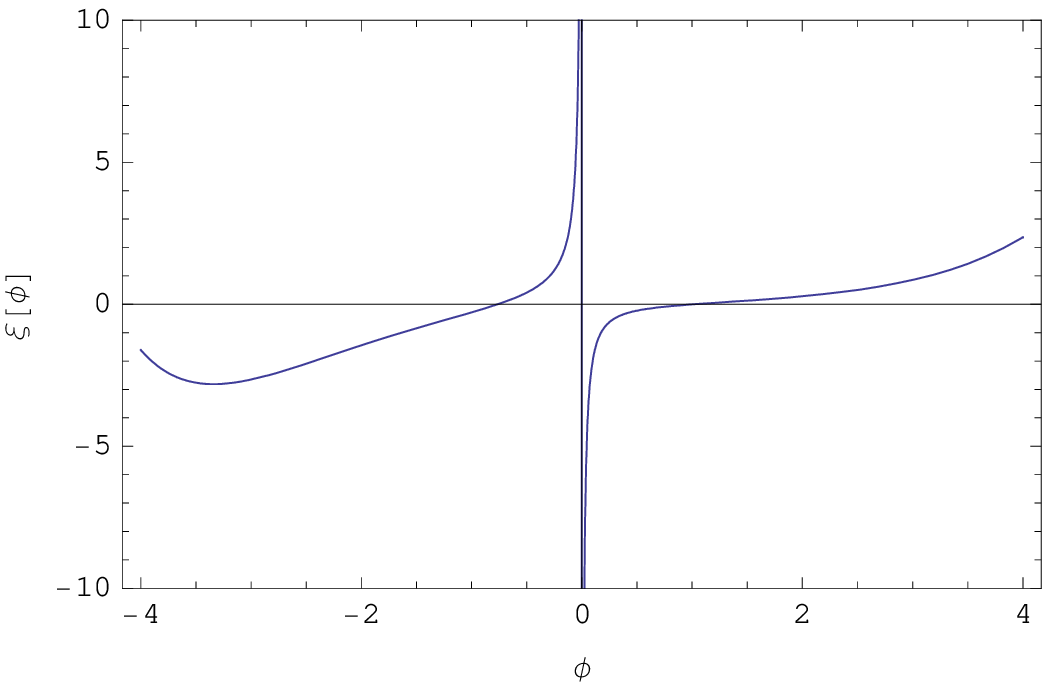}
\end{center}
\caption{$V(\phi)$ (left panel) and $\xi(\phi)$ (right panel)
as functions of $\phi$ for $a(t)=\cosh\lambda t$, $c_1=0$, $c_2=0$, $\lambda=1$, and $\phi_0=1$.
}
\label{graph_3}
\end{figure}

In Fig.~\ref{graph_3}, we display the behaviours of 
$V(\phi)$ and $\xi(\phi)$ 
as functions of $\phi$ for $a(t)=\cosh\lambda t$,
$c_1=0$, $c_2=0$, $\lambda=1$, and $\phi_0=1$.

\subsection{Exponential model}

When the scale factor is described by
$a(t)=\exp\left(\alpha t^2\right)$ with $\alpha>0$ in (\ref{eq:3.15}), 
$V(\phi)$ and $\xi(\phi)$ are reconstructed as
\begin{eqnarray}
V(\phi) \Eqn{=}
\frac{1}{2}\phi_0^2
+\frac{12\alpha^2}{\phi_0^2}\phi^2
+c_1\frac{48\alpha}{\phi_0}\phi \exp\left(\phi+\frac{\alpha\phi^2}{\phi_0^2}\right)
\nonumber \\
&&
{}-\frac{3\sqrt{\pi\alpha}}{\phi_0}(4\alpha-\phi_0^2)
\phi \exp\left(\phi+\frac{\alpha\phi^2}{\phi_0^2}+\frac{\phi_0^2}{4\alpha}
\right)
\erf\left(\frac{\sqrt{\alpha}}{\phi_0}\phi+\frac{\phi_0}{2\sqrt{\alpha}}
\right)\,,
\label{eq:5.13} \\
\xi(\phi) \Eqn{=}
c_2-c_1\left(\frac{\phi_0}{4\alpha^2\phi}
\exp\left(\frac{\alpha\phi^2}{\phi_0^2}\right)
+\frac{\sqrt{\pi}}{4\alpha^{3/2}}\erfi\left(\frac{\sqrt{\alpha}}{\phi_0}\phi\right)\right)
-\frac{\phi_0^2}{32\alpha^2}\left(\frac{1}{\phi}\e^{-\phi}+\ei(-\phi)\right)
\nonumber \\
&&
{}-\frac{\sqrt{\pi}\phi_0}{64\alpha^{5/2}}(4\alpha-\phi_0^2)
\int\frac{1}{\phi^2}
\exp\left(\frac{\alpha}{\phi_0^2}\phi^2+\frac{\phi_0^2}{4\alpha}\right)
\erf\left(\frac{\sqrt{\alpha}}{\phi_0}\phi+\frac{\phi_0}{2\sqrt{\alpha}}\right)d\phi \,,
\label{eq:5.14}
\end{eqnarray}
where $\ei(-\phi)$ is an integration exponential function. 
We also find that
\begin{equation}
V(\phi)\rightarrow \frac{1}{2}\phi_0^2,\quad
\xi(\phi)\rightarrow\infty, \quad
\xi(\phi)\mathcal{G}\rightarrow 0 \
\quad\text{for}\quad \phi\rightarrow 0 \,.
\label{eq:5.15}
\end{equation}
As a special case, by taking $\phi(t)=\pm\sqrt{4\alpha}t$, we have
\begin{eqnarray}
V(\phi) \Eqn{=} 24c_1\sqrt{\alpha}\phi
\exp\left(\phi+\frac{1}{4}\phi^2\right)+2\alpha+3\alpha\phi^2 \,,
\label{eq:5.17} \\
\xi(\phi) \Eqn{=} c_2
-c_1\frac{1}{2\alpha^{3/2}}
\left(\frac{1}{\phi}\exp\left(\frac{\phi^2}{4}\right)
+\frac{\sqrt{\pi}}{2}\erfi\frac{\phi}{2}\right)
-\frac{1}{8\alpha}\frac{1}{\phi}\e^{-\phi}-\frac{1}{8\alpha}\ei(-\phi)\,.
\label{eq:5.18}
\end{eqnarray}
%

\begin{figure}[t]
\begin{center}
\includegraphics[width=7cm,height=5cm]{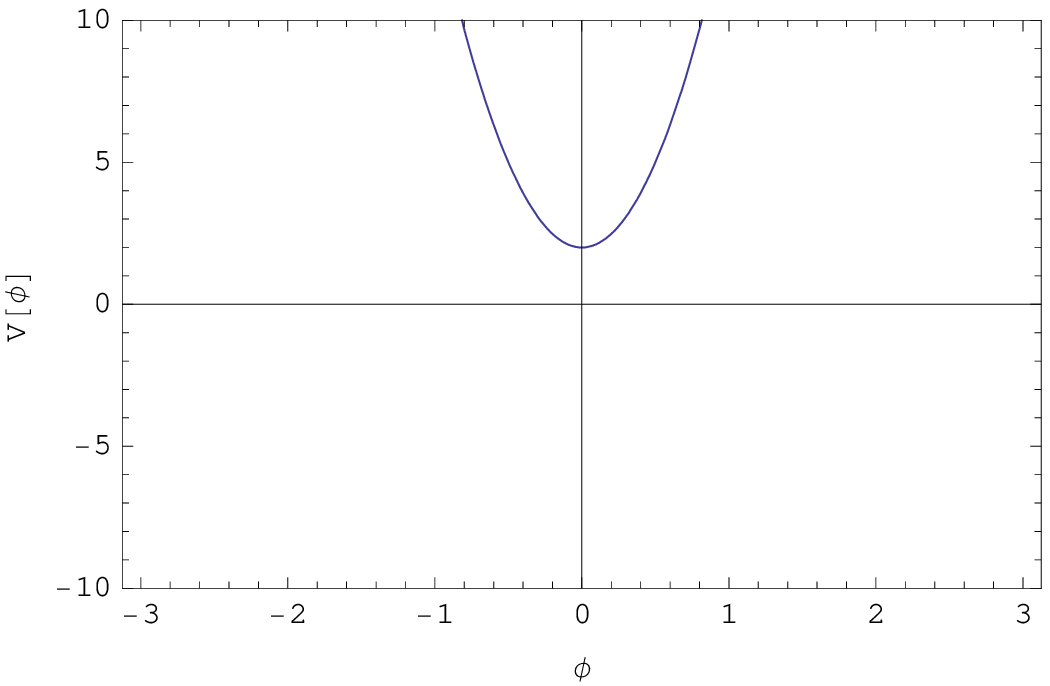}
\hspace{2cm}
\includegraphics[width=7cm,height=5cm]{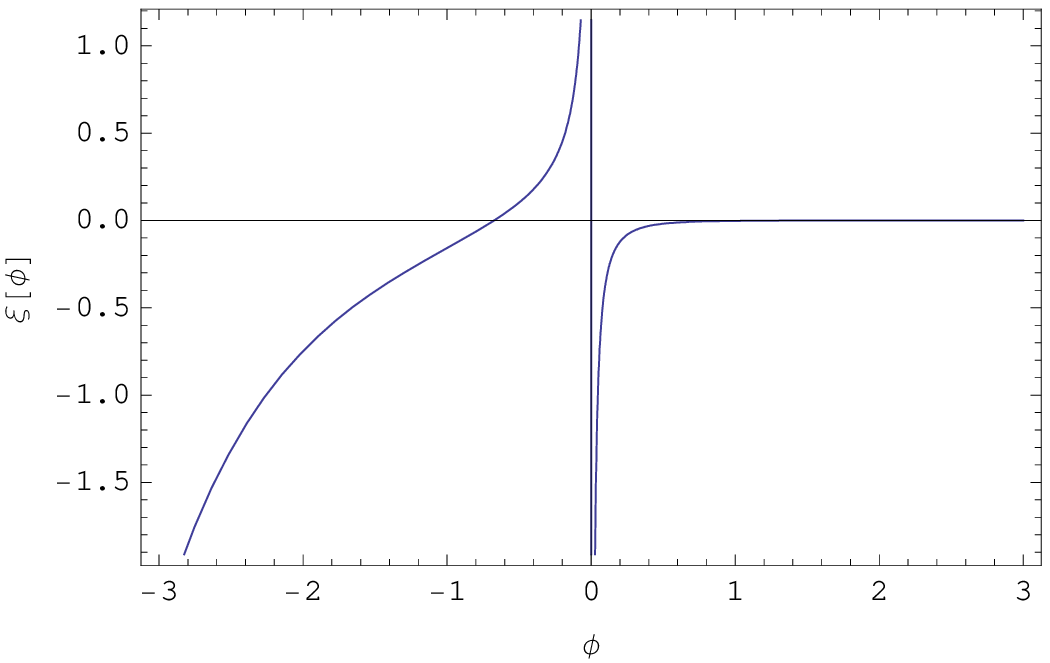}
\end{center}
\caption{$V(\phi)$ (left panel) and $\xi(\phi)$ (right panel)
as functions of $\phi$ for $a(t)=\cosh\lambda t$ with
$\phi=\sqrt{4\alpha}t$, $\phi=\sqrt{4\alpha}t$,
$c_1=0$, $c_2=0$, and $\alpha=1$.}
\label{graph_4}
\end{figure}

In Fig.~\ref{graph_4}, we draw the behaviours of 
$V(\phi)$ and $\xi(\phi)$ 
as functions of $\phi$ for $a(t)=\cosh\lambda t$ with
$\phi=\sqrt{4\alpha}t$, 
$c_1=0$, $c_2=0$, and $\alpha=1$.

\section{Conformal transformation of bounce solutions in the transition from the string frame to the Einstein frame}

In this section, we investigate the transition from the string frame $(g_{\mu\nu}^\mathrm{S},\phi)$ to the Einstein frame $(g_{\mu\nu}^\mathrm{E},\psi)$, 
where $\psi$ is the scalar field in the Einstein frame corresponding to 
the scalar field $\phi$ in the string frame. 
We begin with the action of the heterotic string theory in the string frame
\begin{equation}
S_\mathrm{S}=\int d^4x_\mathrm{S}\sqrt{-g_\mathrm{S}}
\left[\e^{-\phi}\left(\frac{1}{2}R_\mathrm{S}
+\frac{1}{2}(\nabla\phi)^2
-V_\mathrm{S}(\phi)\right)
+\xi_\mathrm{S}(\phi)\mathcal{G}_\mathrm{S}\right]\,.
\label{eq:6.1}
\end{equation}
Here and in the following, the superscription (subscription) ``S'' denotes the quantities in the string frame, whereas the superscription (subscription) ``E'' shows the quantities in the Einstein frame.

We make a conformal transformation~\cite{M-FM-F}
\begin{equation}
g_{\mu\nu}^\mathrm{S}\ \rightarrow\ g_{\mu\nu}^\mathrm{E}=\e^{-\phi}g_{\mu\nu}^\mathrm{S} \,.
\label{eq:6.2}
\end{equation}
The FLRW metric in the string frame is written as 
$ds^2=\e^{-\phi}(-dt^2_\mathrm{S}+a^2_\mathrm{S}dx^2_\mathrm{S})$.
Hence, the relation between time in the Einstein frame and that in the string frame becomes 
\begin{equation}
dt_\mathrm{E}=\pm \e^{-\phi/2}dt_\mathrm{S} \,.
\label{eq:6.3}
\end{equation}
In the further considerations, we choose the positive sign in
Eq.~(\ref{eq:6.3}). This means that the direction of 
motion along the time axis in the string frame 
is the same as that in the Einstein frame. 
There are also the following relations of various quantities 
between in the Einstein and string frames
\begin{eqnarray}
a_\mathrm{E} \Eqn{=} \e^{-\phi/2}a_\mathrm{S} \,,
\label{eq:6.7} \\
H_\mathrm{E} \Eqn{=}
\e^{\phi/2}\left(H_\mathrm{S}-\frac{1}{2}\Dot{\phi}\right) \,,
\label{gb17} \\
\Dot{H}_\mathrm{E} \Eqn{=}
\e^{\phi}\left(\Dot{H}_\mathrm{S}-\frac{1}{2}\Ddot{\phi}+\frac{1}{2}\Dot{\phi}H_\mathrm{S}-\frac{1}{4}\Dot{\phi}^2\right)\,,
\label{eq:6.9} \\
R_\mathrm{E} \Eqn{=}
\e^{\phi}\left(R_\mathrm{S}-9\Dot{\phi}H_\mathrm{S}+\frac{3}{2}\Dot{\phi}^2-3\Ddot{\phi}\right)\,,
\label{eq:6.10} \\
\mathcal{G}_\mathrm{E} \Eqn{=} \e^{2\phi}\left[\mathcal{G}_\mathrm{S}
-3H_\mathrm{S}\left(12H_\mathrm{S}^2\Dot{\phi}-6H_\mathrm{S}\Dot{\phi}^2+\Dot{\phi}^3\right)-\frac{d}{dt}\left(12H_\mathrm{S}^2\Dot{\phi}-6H_\mathrm{S}\Dot{\phi}^2+\Dot{\phi}^3\right)\right]\,.
\label{eq:6.11}
\end{eqnarray}
When the string frame moves to the Einstein frame, 
the potential of the scalar field and its coupling function to the Gauss-Bonnet invariant are changed as follows 
\begin{eqnarray}
V_\mathrm{E}(\psi) \Eqn{=} \e^{\phi}V_\mathrm{S}(\phi)\,,
\label{eq:6.5} \\
\xi_\mathrm{E}(\psi) \Eqn{=} \xi_\mathrm{S}(\phi)\,.
\label{eq:6.6}
\end{eqnarray}

The action in the Einstein frame is given by
\begin{equation}
S_\mathrm{E}=\int d^4x_\mathrm{E}\sqrt{-g_\mathrm{E}}
\left[
\frac{1}{2}R_\mathrm{E}
-\frac{1}{2}(\nabla_E\psi)^2
-V_\mathrm{E}(\psi)
+\xi_\mathrm{E}(\psi)\left(\mathcal{G}_\mathrm{E}+\mathcal{F}(\nabla_\mathrm{E}\psi,R_\mathrm{E})\right)
\right]\,,
\label{eq:v5-6-11}
\end{equation}
where $g_\mathrm{E}$ is the determinant of the metric $g_{\mu\nu}^\mathrm{E}$ in the Einstein frame, $R_\mathrm{E}$ is the Ricci scalar, and $\mathcal{G}_\mathrm{E}$ are the Gauss-Bonnet invariant. 
In this action, there is the Gauss-Bonnet term. 
Hence, if we start from the effective action in the string frame, 
the additional term $\mathcal{F}$ appears in the Einstein frame~\cite{Maeda:2009}. 

In the Einstein frame, the conditions for the existence of 
the cosmological bounce at $t_b ^\mathrm{E}$ are 
$H_\mathrm{E}(t_\mathrm{b}^\mathrm{E})=0$ and $\Dot{H}_\mathrm{E}(t_\mathrm{b}^\mathrm{E})>0$. These relations have to be fulfilled at the bounce point. 
The corresponding conditions in the string frame to these relations 
in the Einstein frame are represented as 
\begin{equation}
H_\mathrm{S}-\frac{1}{2}\Dot{\phi}=0\,,
\quad
\Dot{H}_\mathrm{S}+\frac{1}{2}\Dot{\phi}H_\mathrm{S}-\frac{1}{4}\Dot{\phi}^2-\frac{1}{2}\Ddot{\phi}>0\,.
\label{gb20}
\end{equation}

If the scalar field linearly dependents on time as 
$\phi(t_\mathrm{S})=\phi_0 t_\mathrm{S}$, by taking Eq.~(\ref{eq:6.3}) into consideration, we get
\begin{equation}
t_\mathrm{E}=-\frac{2}{\phi_0}\exp\left(-\frac{1}{2}\phi_0t_\mathrm{S}\right)
\quad\rightarrow\quad
t_\mathrm{S}=-\frac{2}{\phi_0}\ln\left(-\frac{1}{2}\phi_0t_\mathrm{E}\right),\quad
t_\mathrm{E}\phi_0<0\,.
\label{eq:v5-6-13}
\end{equation}
Under the conformal transformation, 
the time axis $t_\mathrm{S}$ converts into a positive or negative time 
semi-axis $t_\mathrm{E}$. Furthermore, if $\phi_0>0 \, (<0)$, 
the mapping occurs on the negative (positive) 
semi-axis $t_\mathrm{E}$. 

We analyze the behaviours of the scale factor 
$a_\mathrm{E}(t_\mathrm{E})$ and its second derivative 
$\Ddot{a}_\mathrm{E}(t_\mathrm{E})$ around 
$t_\mathrm{E}^\star$, which corresponds to the point of the 
cosmological bounce $t_\mathrm{b}^\mathrm{S}$ in the string frame. 
It is known that $\Dot{a}_\mathrm{S}(t_\mathrm{b}^\mathrm{S})=0$, 
and therefore the scale factor at $t_\mathrm{b}^\mathrm{S}$ has an extreme value. 
Using the relation $\phi(t_\mathrm{S})=\phi_0t_\mathrm{S}$ and 
Eq.~(\ref{eq:6.7}), we can determine the values of higher derivatives of the scale factor at the point of $t_\mathrm{E}^\star$ in the Einstein frame as 
\begin{eqnarray}
\Ddot{a}_\mathrm{E}(t^\star_\mathrm{E}) \Eqn{=}
\exp\left(\frac{1}{2}\phi_0t_\mathrm{b}^\mathrm{S}\right)\Ddot{a}_\mathrm{S}(t_\mathrm{b}^\mathrm{S})\,,
\label{eq:6.32} \\
a_\mathrm{E}^{(3)}(t^\star_\mathrm{E}) \Eqn{=}
\exp\left(\phi_0t_\mathrm{b}^\mathrm{S}\right)a_\mathrm{S}^{(3)}
(t_\mathrm{b}^\mathrm{S})\,,
\label{eq:6.33} \\
a_\mathrm{E}^{(4)}(t^\star_\mathrm{E}) \Eqn{=}
\exp\left(\frac{3}{2}\phi_0t_\mathrm{b}^\mathrm{S}\right)
\left(a_\mathrm{S}^{(4)}(t_\mathrm{b}^\mathrm{S})
-\frac{1}{4}\phi_0^2\Ddot{a}_\mathrm{S}(t_\mathrm{b}^\mathrm{S})+\phi_0a_\mathrm{S}^{(3)}(t_\mathrm{b}^\mathrm{S})\right)\,.
\label{eq:6.34}
\end{eqnarray}
It follows that the sign of second and third derivatives of the 
scale factor around $t_\mathrm{b}^\mathrm{S}$ will be 
maintained during the transition from the string frame to the 
Einstein frame. However, if the function 
$\Ddot{a}_\mathrm{S}(t_\mathrm{S})$ has an extreme value at 
$t_\mathrm{b}^\mathrm{S}$, the function 
$\Ddot{a}_\mathrm{E}(t_\mathrm{E})$ will also have an extreme value, 
but it has already had an extreme value at the point $t_\mathrm{E}^\star$.


\subsection{Hyperbolic model in the string frame}

We study the case that the scale factor in the string frame
has the hyperbolic form
\begin{equation}
a_\mathrm{S}(t_\mathrm{S})=\sigma \e^{\lambda t_\mathrm{S}}+\tau \e^{-\lambda t_\mathrm{S}}\,,
\quad
\lambda>0\,.
\label{eq:6.12}
\end{equation}
In this model, the conditions in (\ref{gb20}) for the existence of the cosmological bounce read
\begin{equation}
|\phi_0|<2\lambda \,,
\quad
\sigma\tau>0 \,.
\label{eq:v5-6-18}
\end{equation}
{}From Eqs.~(\ref{eq:6.7}) and (\ref{gb17}), we have
\begin{eqnarray}
a_\mathrm{E}(t_\mathrm{E}) \Eqn{=} \frac{2^{\frac{4\lambda}{\phi_0}}\sigma+(-\phi_0t_\mathrm{E})^{\frac{4\lambda}{\phi_0}}\tau}
{4(-2\phi_0t_\mathrm{E})^{\frac{2\lambda}{\phi_0}-1}}\,,
\quad \phi_0t_\mathrm{E}<0 \,,
\label{eq:v5-6-19} \\
H_\mathrm{E}(t_\mathrm{E})\Eqn{=}
\frac{1}{t_\mathrm{E}}+\frac{2\lambda}{t_\mathrm{E}\phi_0}
\left[
1-\frac{2^{1+\frac{4\lambda}{\phi_0}}\sigma}
{2^{\frac{4\lambda}{\phi_0}}\sigma+(-\phi_0t_\mathrm{E})^{\frac{4\lambda}{\phi_0}}\tau}
\right]\,,
\quad
\phi_0t_\mathrm{E}<0\,.
\label{eq:v5-6-20}
\end{eqnarray}
We define the point of the cosmological bounce in the Einstein frame as 
\begin{equation}
H_\mathrm{E}(t_\mathrm{E})=0\quad\rightarrow\quad
t_\mathrm{b}^\mathrm{E} \equiv -\frac{2}{\phi_0}
\left(\frac{\sigma}{\tau}\right)^{\frac{\phi_0}{4\lambda}}
\left(\frac{2\lambda-\phi_0}{2\lambda+\phi_0}\right)^{\frac{\phi_0}{4\lambda}},
\quad \phi_0t_\mathrm{E}<0\,.
\label{eq:v5-6-21}
\end{equation}
For the model in the string frame, under the conformal transformation, 
the point of the cosmological bounce $t_\mathrm{b}^\mathrm{S}$ moves to a point $t^\star_\mathrm{E}$ in the Einstein frame 
\begin{equation}
t_\mathrm{b}^\mathrm{S}=\frac{1}{2\lambda}\ln\left(\frac{\tau}{\sigma}\right)
\quad\rightarrow\quad
t^\star_\mathrm{E}=-\frac{2}{\phi_0}\left(\frac{\sigma}{\tau}\right)^{\frac{\phi_0}{4\lambda}}\,,
\quad \phi_0t_\mathrm{E}<0\,.
\label{eq:v5-6-22}
\end{equation}

We examine the behaviour of second derivative of the
scale factor in the Einstein frame. 
By using Eqs.~(\ref{eq:6.33}) and (\ref{eq:6.34}) and taking into account the conditions for the point of the cosmological bounce in the Einstein frame, 
we obtain
\begin{equation}
a_\mathrm{E}^{(3)}(t^\star_\mathrm{E})=0\,,
\quad
a_\mathrm{E}^{(4)}(t^\star_\mathrm{E})=
\frac{1}{2}\lambda^2\sigma(4\lambda^2-\phi_0^2)
\left(\frac{\tau}{\sigma}\right)^{\frac{1}{2}+\frac{3\phi_0}{4\lambda}}>0\,.
\label{eq:v5-6-23}
\end{equation}
Accordingly, it is seen that at the point $t^\star_\mathrm{E}$, 
$\Ddot{a}_\mathrm{E}(t_\mathrm{E})$ has a minimum 
\begin{equation}
(a_\mathrm{E})_\mathrm{min}=a_\mathrm{E}(t_\mathrm{E}^\star)=
2\lambda^2\sqrt{\sigma\tau}\left(\frac{\tau}{\sigma}\right)^{\frac{\phi_0}{4\lambda}}\,.
\label{eq:v5-6-24}
\end{equation}
It can be shown that the function has only this extreme value. 

\begin{figure}[t]
\begin{center}
\includegraphics[width=7cm,height=5cm]{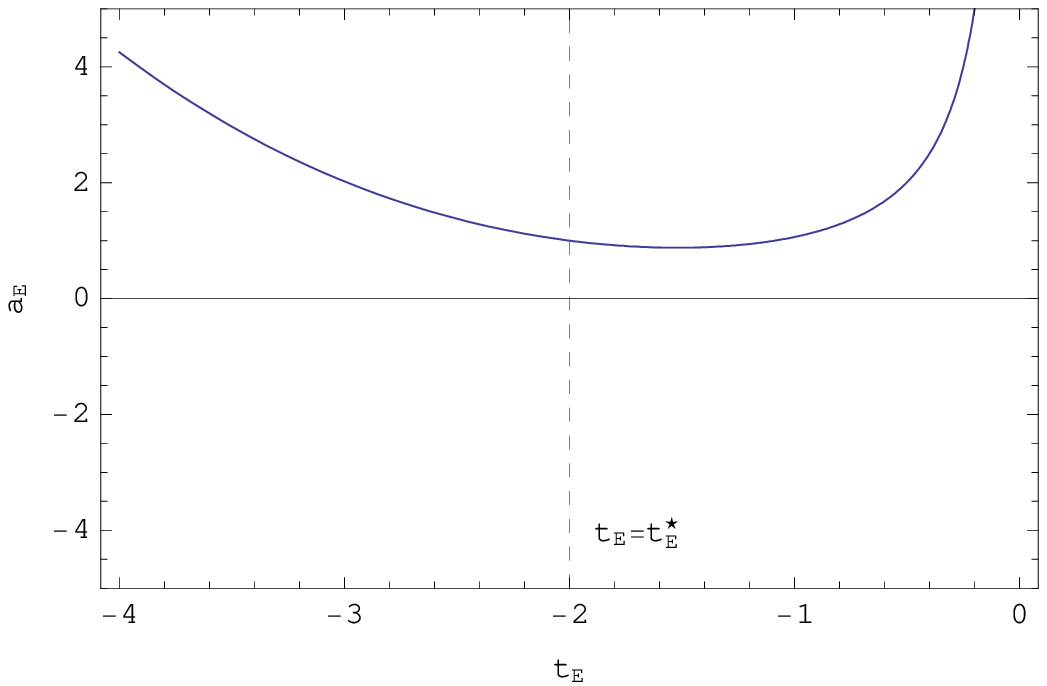}
\hfill
\includegraphics[width=7cm,height=5cm]{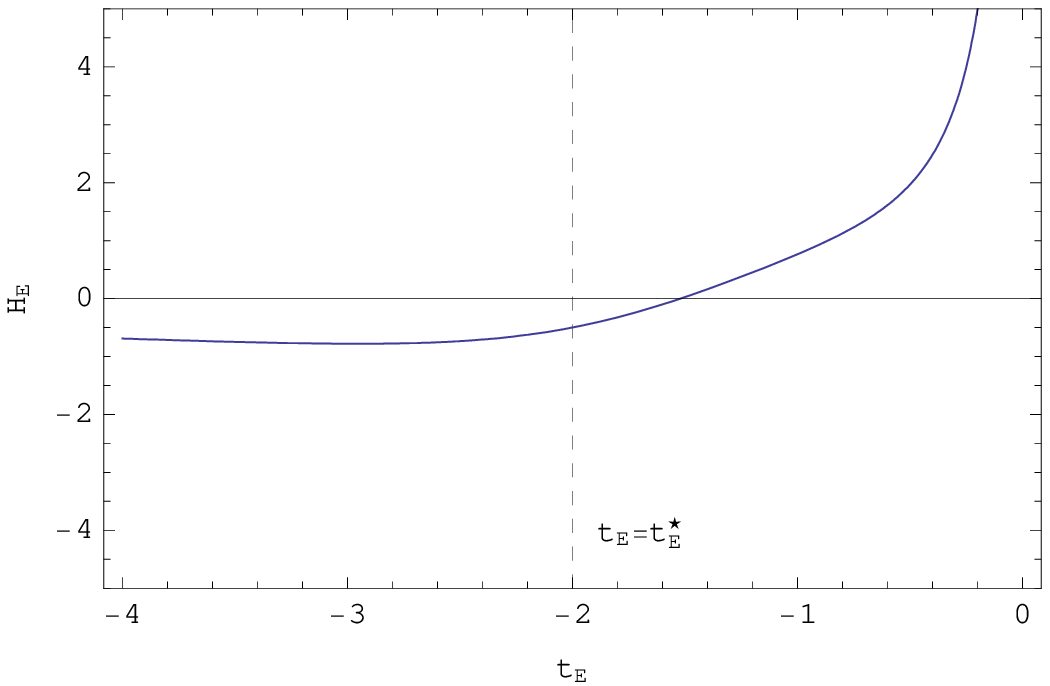}
\end{center}
\caption{$a_\mathrm{E}(t_\mathrm{E})$ (left panel) and
$H_\mathrm{E}(t_\mathrm{E})$ (right panel)
as functions of $t_\mathrm{E}$
for $a_\mathrm{S}(t_\mathrm{S})=\cosh \left(\lambda t_\mathrm{S}\right)$ with
$\phi(t_\mathrm{S})=\lambda t_\mathrm{S}$ and $\lambda=1$.
The dashed line shows 
$t_\mathrm{E}=-2/\lambda$ corresponding to the time of 
the cosmological bounce in the string frame.}
\label{graph_5}
\end{figure}

We explore the special case that 
$\sigma=\tau=1/2$ and $\phi_0=\lambda$. 
In this case, we find 
\begin{equation}
a_\mathrm{S}(t_\mathrm{S})=\cosh \left(\lambda t_\mathrm{S}\right)\,,
\quad
\phi(t_\mathrm{S})=\lambda t_\mathrm{S},\quad \lambda>0\,.
\label{eq:v5-6-25}
\end{equation}
In Fig.~\ref{graph_5}, we show the evolutions of
$a_\mathrm{E}(t_\mathrm{E})$ and $H_\mathrm{E}(t_\mathrm{E})$
as functions of $t_\mathrm{E}$
for $a_\mathrm{S}(t_\mathrm{S})=\cosh \left(\lambda t_\mathrm{S}\right)$ with
$\phi(t_\mathrm{S})=\lambda t_\mathrm{S}$ and $\lambda=1$. 

\begin{figure}[t]
\includegraphics[width=7cm,height=5cm]{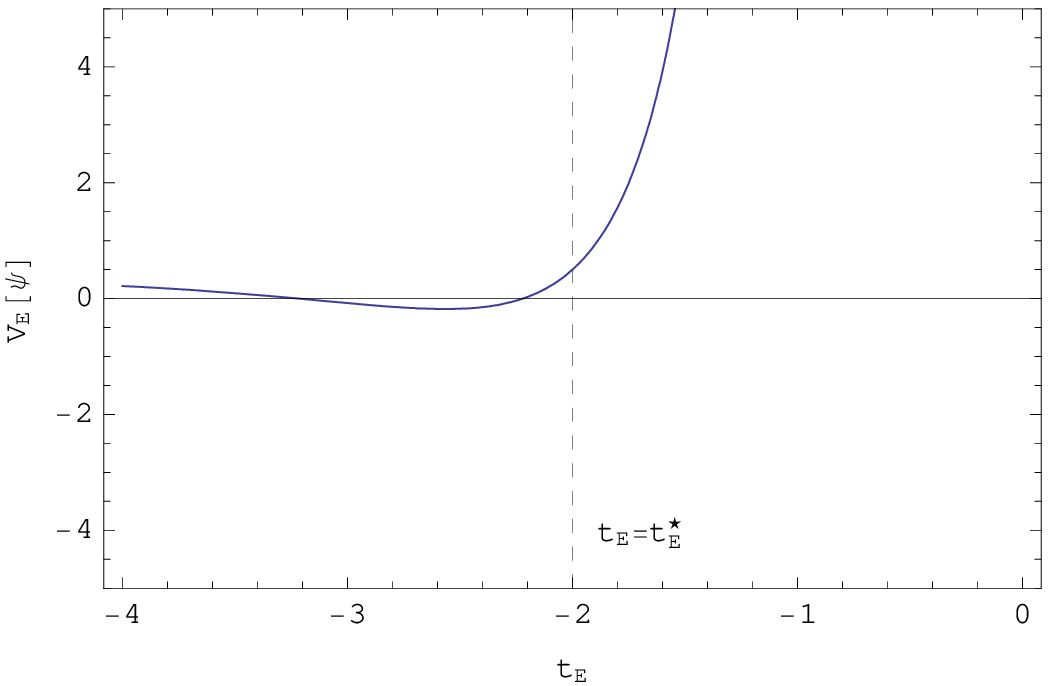}
\hfill
\includegraphics[width=7cm,height=5cm]{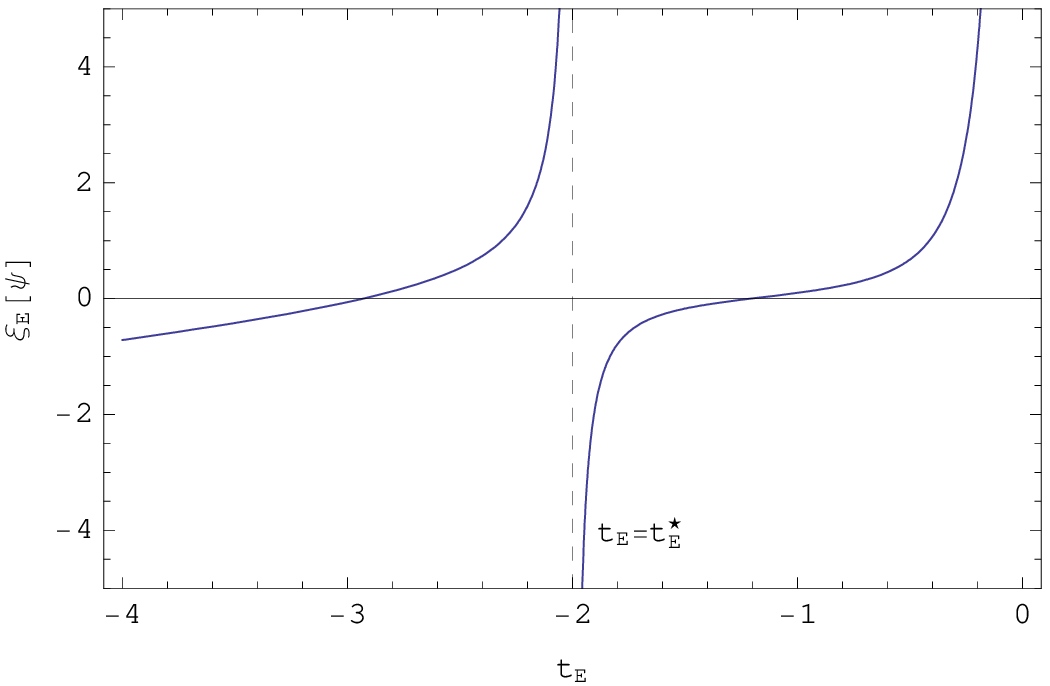}
\\
\caption{$V_\mathrm{E}(t_\mathrm{E})$ and
$\xi_\mathrm{E}(t_\mathrm{E})$ as functions of $t_\mathrm{E}$ for
$a_\mathrm{S}(t_\mathrm{S})=\cosh \left(\lambda t_\mathrm{S}\right)$ with
$\phi(t_\mathrm{S})=\lambda t_\mathrm{S}$, $c_1=0$, $c_2=0$, and
$\lambda=1$. Legend is the same as Fig.~\ref{graph_5} with
$t_\mathrm{E}=-2/\lambda$.}
\label{graph_6}
\end{figure}

The functions $V_\mathrm{E}(\psi)$ and $\xi_\mathrm{E}(\psi)$
are reconstructed as follows 
\begin{eqnarray}
V_\mathrm{E}(\psi(t_\mathrm{E})) \Eqn{=}
c_1\frac{48(16-t_\mathrm{E}^4\lambda^4)}{t_\mathrm{E}^6\lambda^5}
+\frac{224-10t_\mathrm{E}^4\lambda^4}{t_\mathrm{E}^2
(16+t_\mathrm{E}^4\lambda^4)}
\nonumber \\
&&
{}-\frac{6(16-t_\mathrm{E}^4\lambda^4)}{t_\mathrm{E}^6\lambda^4}
\ln\left(\frac{1}{32}(16+t^4_\mathrm{E}\lambda^4)\right), \quad
t_\mathrm{E}<0 \,.
\label{eq:6.19} \\
\xi_\mathrm{E}(\psi(t_\mathrm{E})) \Eqn{=}
c_2+c_1\frac{256-96\lambda^4t_\mathrm{E}^4+\lambda^8t_\mathrm{E}^8}{8\lambda^5t_\mathrm{E}^2(16-\lambda^4t_\mathrm{E}^4)}
-(4+\frac{15}{2}\ln
2)\frac{t_\mathrm{E}^2}{16-\lambda^4t_\mathrm{E}^4}
\nonumber \\
&&
{}+\frac{1}{2\lambda^2}\arccot\frac{4}{\lambda^2t_\mathrm{E}^2}
+\frac{1}{2\lambda^2}\arccoth
\left(\exp\left(\left|\ln(\lambda^2t_\mathrm{E}^2)-2\ln 2\right|\right)\right)
\nonumber \\
&&
{}+\frac{t_\mathrm{E}^2}{2(16-\lambda^4t_\mathrm{E}^4)}
\left[
\left(3-\frac{8}{\lambda^4t_\mathrm{E}^4}-\frac{1}{32}\lambda^4t_\mathrm{E}^4\right)\ln\left(1+\frac{16}{\lambda^4t_\mathrm{E}^4}\right)
\right.
\nonumber \\
&&
\left.
{}+\left(\frac{8}{\lambda^4t_\mathrm{E}^4}+\frac{1}{32}\lambda^4t_\mathrm{E}^4\right) \left(4+5\ln 2-2\ln(\lambda^2t_\mathrm{E}^2)\right)
\right.
\nonumber \\
&& \left.
{}+6\ln(\lambda^2t_\mathrm{E}^2)-\frac{32}{\lambda^4t_\mathrm{E}^4}+\frac{1}{8}\lambda^4t_\mathrm{E}^4
\right], \quad t_\mathrm{E}<0 \,.
\label{eq:6.21}
\end{eqnarray}
In Fig.~\ref{graph_6}, we illustrate the evolutions of 
$V_\mathrm{E}(t_\mathrm{E})$ and $\xi_\mathrm{E}(t_\mathrm{E})$
as functions of $t_\mathrm{E}$ 
for $a_\mathrm{S}(t_\mathrm{S})=\cosh \left(\lambda t_\mathrm{S}\right)$ with
$\phi(t_\mathrm{S})=\lambda t_\mathrm{S}$,
$c_1=0$, $c_2=0$, and $\lambda=1$.


\subsection{Exponential model in the string frame}

We discuss the case that the scale factor in the string frame is
written as
\begin{equation}
a_\mathrm{S}(t_\mathrm{S})=\exp\left(\alpha t_\mathrm{S}^2\right)\,,
\quad
\alpha>0\,,
\label{eq:6.22}
\end{equation}
In this model, the point of the cosmological bounce in the Einstein
frame exists for any non-zero values of the parameter $\phi_0$.
With Eqs.~(\ref{eq:6.7}) and (\ref{gb17}), we acquire
\begin{eqnarray}
a_\mathrm{E}(t_\mathrm{E}) \Eqn{=}
-\frac{1}{2}\phi_0t_\mathrm{E}\exp \left( \frac{4\alpha}{\phi^2_0}
\ln^2 \left(-\frac{1}{2}\phi_0t_\mathrm{E}\right)\right)\,,
\quad \phi_0t_\mathrm{E}<0\,,
\label{eq:v5-6-29} \\
H_\mathrm{E}(t_\mathrm{E}) \Eqn{=}
\frac{1}{t_\mathrm{E}}+\frac{8\alpha\ln\left(-\frac{1}{2}\phi_0t_\mathrm{E}\right)}{\phi_0^2t_\mathrm{E}},\quad \phi_0t_\mathrm{E}<0\,.
\label{eq:v5-6-30}
\end{eqnarray}
We define the point of the cosmological bounce in the Einstein frame as 
\begin{equation}
H_\mathrm{E}(t_\mathrm{E})=0\quad\rightarrow\quad
t_\mathrm{b}^\mathrm{E} \equiv -\frac{2}{\phi_0}\exp\left(-\frac{\phi_0^2}{8\alpha}\right)\,,
\quad
\phi_0t_\mathrm{E}<0.
\label{eq:v5-6-31}
\end{equation}
For this model in Eq.~(\ref{eq:6.22}), 
through the conformal transformation, 
the point of the cosmological bounce $t_\mathrm{b}^\mathrm{S}$ in the string frame moves to a point $t^\star_\mathrm{E}$ in the Einstein frame 
\begin{equation}
t_\mathrm{b}^\mathrm{S}=0\quad\rightarrow\quad
t^\star_\mathrm{E}=-\frac{2}{\phi_0}\,, \quad
\phi_0t_\mathrm{E}<0\,.
\label{eq:v5-6-32}
\end{equation}
%

\begin{figure}[t]
\begin{center}
\includegraphics[width=7cm,height=5cm]{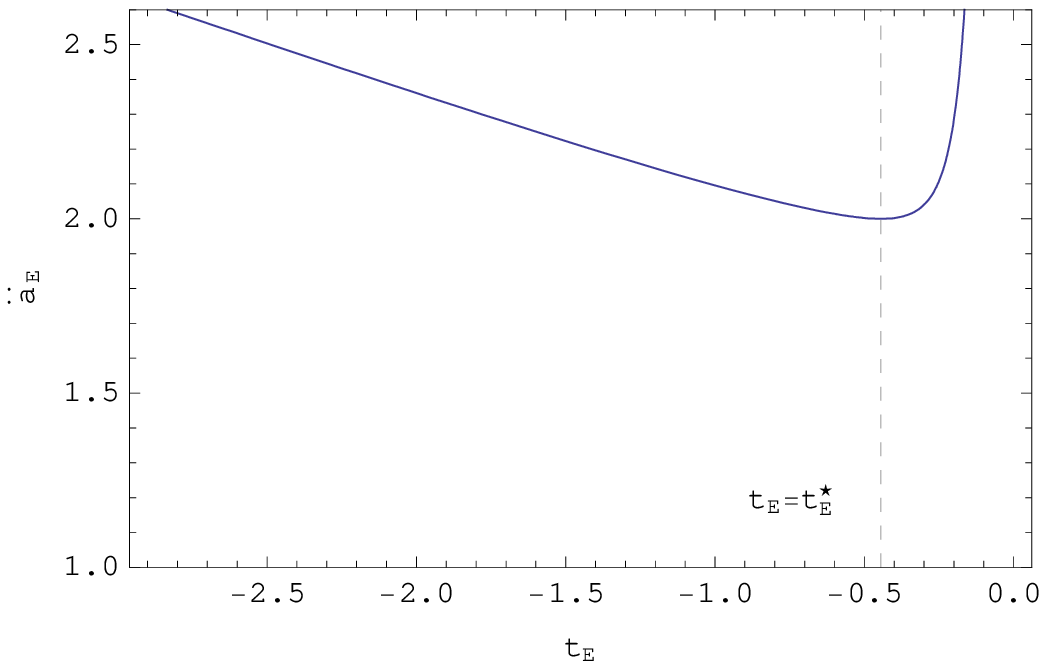}
\hfill
\includegraphics[width=7cm,height=5cm]{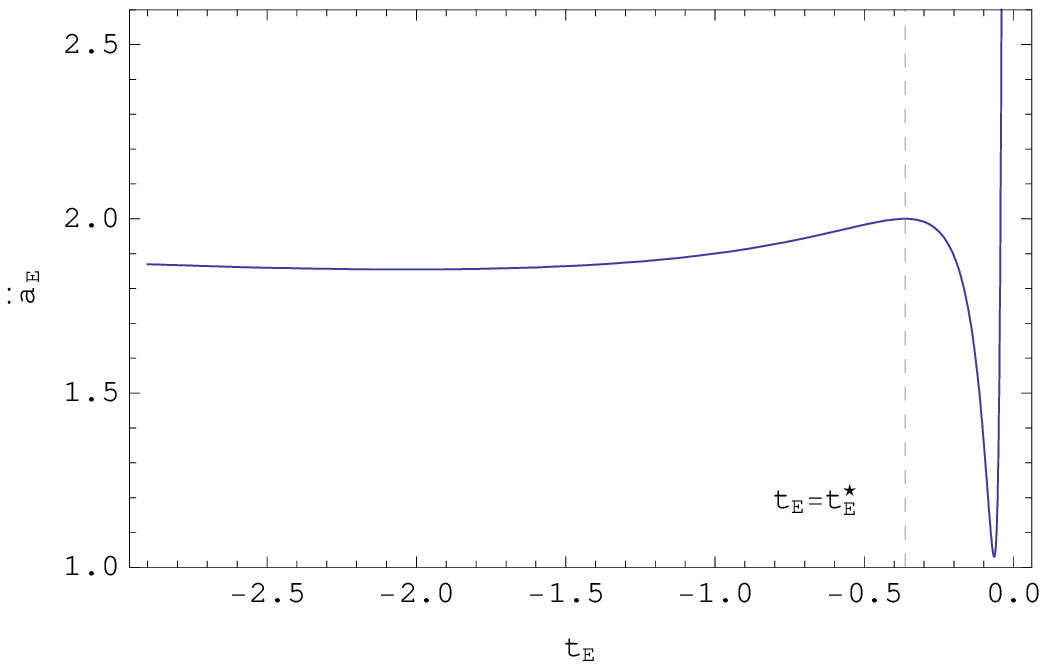}
\end{center}
\caption{$\Ddot{a}_\mathrm{E}(t_\mathrm{E})$ for
$a_\mathrm{S}(t_\mathrm{S})=\exp\left(\alpha t_\mathrm{S}^2\right)$ 
with $\alpha=1$, $\phi_0=4.5$ (left panel) or $\phi_0=5.5$ (right panel).
The dashed line intersects the time axis $t_\mathrm{E}$ at
$t_\mathrm{E}=t_\mathrm{E}^\star$, which corresponds to the
cosmological bounce time in the string frame.}
\label{graph_7}
\end{figure}

We also study the behaviour of the function 
$\Ddot{a}_\mathrm{E}(t_\mathrm{E})$ for this model. 
It follows from Eqs.~(\ref{eq:6.33}) and (\ref{eq:6.34}) 
that at the point $t_\mathrm{E}^\star$, the function has an extreme value. 
Hence, when $|\phi_0|< \sqrt{24\alpha} \,\, (|\phi_0| > \sqrt{24\alpha})$, 
it has a minimum (maximum). 
Moreover, from Fig.~\ref{graph_7}, it is found that 
in the latter case of $|\phi_0| > \sqrt{24\alpha}$, 
the function $\Ddot{a}_\mathrm{E}(t_\mathrm{E})$ has an additional 
extreme value at the points 
\begin{equation}
t_\mathrm{E}^{(-)}=-\frac{2}{\phi_0}\exp\left(\frac{|\phi_0|}{8\alpha}\sqrt{\phi_0^2-24\alpha}\right)\,,
\quad
t_\mathrm{E}^{(+)}=-\frac{2}{\phi_0}\exp\left(-\frac{|\phi_0|}{8\alpha}\sqrt{\phi_0^2-24\alpha}\right)\,,
\quad \phi_0t_\mathrm{E}<0 \,.
\label{eq:v5-6-33}
\end{equation}
In this case, we get
\begin{eqnarray}
(a_\mathrm{E})_\mathrm{max} \Eqn{=}
a_\mathrm{E}(t_\mathrm{E}^\star)=2\alpha \,,
\label{eq:v5-6-34} \\
(a_\mathrm{E})_\mathrm{min} \Eqn{=}
a_\mathrm{E}(t_\mathrm{E}^{(\mp)}) =
\frac{1}{4}\exp\left(\frac{-24\alpha+\phi_0^2}{16\alpha}\mp\frac{|\phi_0|\sqrt{-24\alpha+\phi_0^2}}{8\alpha}\right)
\nonumber \\
&&
\hspace{25mm}
{}\times \left(-16\alpha+\phi_0^2\pm|\phi_0|\sqrt{-24\alpha+\phi_0^2}\right)\,,
\quad
\phi_0t_\mathrm{E}<0 \,.
\label{eq:v5-6-35}
\end{eqnarray}
%

\begin{figure}[t]
\begin{center}
\includegraphics[width=7cm,height=5cm]{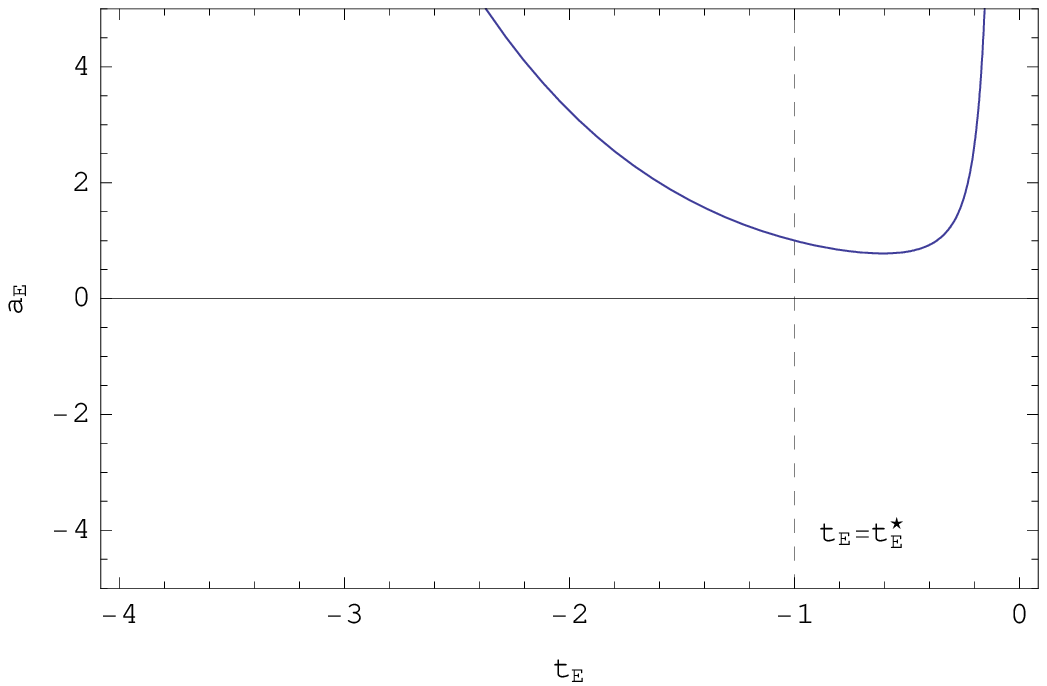}
\hfill
\includegraphics[width=7cm,height=5cm]{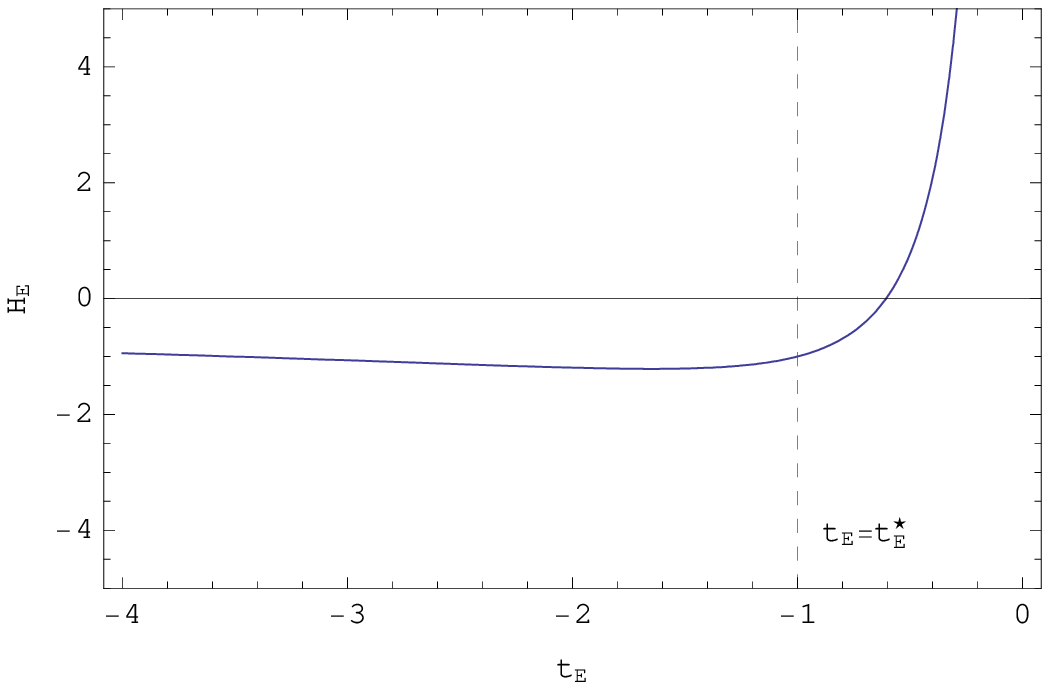}
\end{center}
\caption{$a_\mathrm{E}(t_\mathrm{E})$ (left panel) and
$H_\mathrm{E}(t_\mathrm{E})$ (right panel) as functions of $t_\mathrm{E}$ for
$a_\mathrm{S}(t_\mathrm{S})=\exp\left(\alpha t_\mathrm{S}^2\right)$ with
$\phi(t_\mathrm{S})=\sqrt{4\alpha}t_\mathrm{S}$ and $\alpha=1$.
The dashed line shows 
$t_\mathrm{E}=-1/\sqrt{\alpha}$, which corresponds to the time 
of the cosmological bounce in the string frame.}
\label{graph_8}
\end{figure}

\begin{figure}[t]
\begin{center}
\includegraphics[width=7cm,height=5cm]{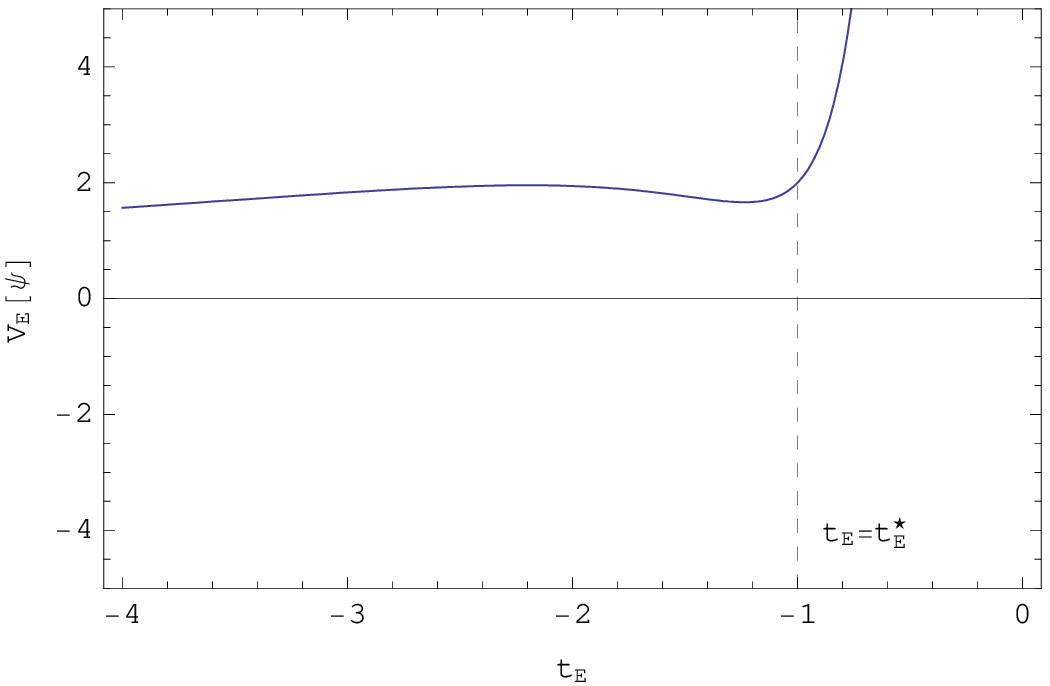}
\hfill
\includegraphics[width=7cm,height=5cm]{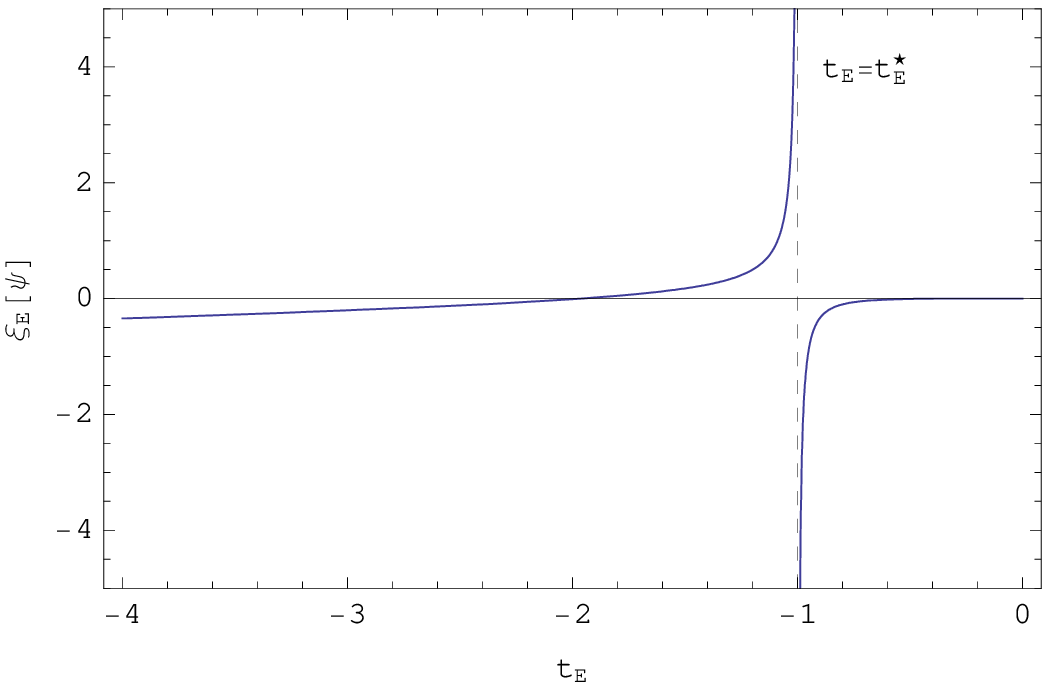}
\end{center}
\caption{$V_\mathrm{E}(t_\mathrm{E})$ (left panel) and
$\xi_\mathrm{E}(t_\mathrm{E})$ (right panel)
as functions of $t_\mathrm{E}$ for
$a_\mathrm{S}(t_\mathrm{S})=\exp\left(\alpha t_\mathrm{S}^2\right)$ with
$\phi(t_\mathrm{S})=\sqrt{4\alpha}t_\mathrm{S}$,
$c_1=0$, $c_2=0$, and $\alpha=1$.
Legend is the same as Fig.~\ref{graph_8}.
}
\label{graph_9}
\end{figure}

Let us consider the case that 
$\phi(t_\mathrm{S})=\phi_0t_\mathrm{S}$ with
$\phi_0=\sqrt{4\alpha}$.  
In Fig.~\ref{graph_8}, we depict the evolutions of 
$a_\mathrm{E}(t_\mathrm{E})$ and $H_\mathrm{E}(t_\mathrm{E})$ 
as functions of $t_\mathrm{E}$ for 
$a_\mathrm{S}(t_\mathrm{S})=\e^{\alpha t_\mathrm{S}^2}$ with
$\phi(t_\mathrm{S})=\sqrt{4\alpha}t_\mathrm{S}$ and $\alpha=1$. 
Furthermore, 
the functions $V_\mathrm{E}(\psi)$ and $\xi_\mathrm{E}(\psi)$ 
are expressed as 
\begin{eqnarray}
V_\mathrm{E}(\psi(t_\mathrm{E})) \Eqn{=}
-c_1\frac{48}{\alpha^{3/2}t_\mathrm{E}^4}
\exp\left(\ln^2\left(-t_\mathrm{E}\sqrt{\alpha}\right)\right)\ln\left(-t_\mathrm{E}\sqrt{\alpha}\right)
+\frac{2}{t_\mathrm{E}^2}
+\frac{12\ln^2\left(-t_\mathrm{E}\sqrt{\alpha}\right)}{t_\mathrm{E}^2}\,, 
\label{eq:6.29} \\
\xi_\mathrm{E}(\psi(t_\mathrm{E})) \Eqn{=}c_2
+c_1\frac{1}{2\alpha^{3/2}}\left(\frac{\sqrt{\pi}}{2}\erfi\left(\ln\left(-t_\mathrm{E}\sqrt{\alpha}\right)\right)
+\frac{\e^{\ln^2\left(-t_\mathrm{E}\sqrt{\alpha}\right)}}{2\ln\left(-t_\mathrm{E}\sqrt{\alpha}\right)}\right)
\nonumber \\
&&
{}+\frac{t_\mathrm{E}^2}{16\ln\left(-t_\mathrm{E}\sqrt{\alpha}\right)}
-\frac{1}{8\alpha} \mathrm{Ei} \left(2\ln\left(-t_\mathrm{E}\sqrt{\alpha}\right)\right)\,, 
\label{eq:6.31}
\end{eqnarray}
where $t_\mathrm{E}<0$. In Fig.~\ref{graph_9}, we plot 
the evolutions of $V_\mathrm{E}(t_\mathrm{E})$ and 
$\xi_\mathrm{E}(t_\mathrm{E})$ 
as functions of $t_\mathrm{E}$ for
$a_\mathrm{S}(t_\mathrm{S})=\e^{\alpha t_\mathrm{S}^2}$ with
$\phi(t_\mathrm{S})=\sqrt{4\alpha}t_\mathrm{S}$,
$c_1=0$, $c_2=0$, and $\alpha=1$.

\section{Conclusions}

In the present paper, we have studied the bounce universe in the framework of scalar Gauss-Bonnet gravity. 
The existence of the Gauss-Bonnet invariant as a higher derivative
quantum correction is strongly supported by string theories. 
Particularly, when the scale factor has the hyperbolic form or 
exponential form leading to cosmology with a bounce, 
we have explicitly reconstructed 
the potential form and Gauss-Bonnet coupling function of a dynamical scalar field. 

In addition, we have explored the bounce behaviours in both the string and Einstein frames in detail by performing the conformal transformation 
and derived the relation of the bounce cosmology 
between these conformal frames. 
Through the conformal transformation, it has been seen that 
the difference of 
potential form of the scalar field 
between the two frames is the exponential function 
of the scalar field, whereas the coupling function 
of the scalar field to the Gauss-Bonnet invariant are the same 
in the two frames. 

As a consequence, we have found the following three points. 
(i) 
In the case that the point of the cosmological bounce in the string frame is transformed into that in the Einstein frame, it does not retain its character. 
When the conformal transformation from the string frame 
to the Einstein frame is made, the bounce point in the string frame 
changes its qualitative natures and ceases to be bounce point in the Einstein frame. However, new bounce point(s) appears in the Einstein frame. 
(ii) 
If the second derivative of the scale factor takes an extreme value 
in the string frame, the second derivative of the scale factor in the Einstein frame has an extreme value at the point corresponding to the one of the cosmological bounce in the string frame. 
Especially, there are cosmological models in which at this point, the 
universe expands with its minimal acceleration, 
namely, the second derivative of the scale factor becomes its minimum value. 
However, in principle, the parameters of the theory may be chosen 
so that the second derivative of the scale factor at this point 
can take its maximum value. 
(iii) 
Third, in the Einstein frame, at the point of the cosmological bounce 
$t_\mathrm{b}^\mathrm{E}$, the gap interaction function $\xi(\phi)$ 
is missed unlike in the string frame. 

In inflation paradigm, 
the spatially flat, homogeneous, and isotropic universe,  
which are suggested by quite precise cosmological observations, 
can be realized successfully. 
The primordial density perturbations 
with its spectrum consistent with the observations 
can also be generated during inflation. 
It is significant to discuss other possible scenarios for the early universe to explain the observations so that physics in the early universe 
can further be proved. 
As an attempt for this issue, 
the idea of the bounce universe has been examined. 
For instance, also in the matter bounce scenario and the Ekpyrotic cosmology, 
the primordial density perturbations 
with its almost scale-invariant spectrum can be generated. 
Another additional merit of these scenarios is that 
they are motivated by fundamental theories including 
superstring/M-theories, which are hopeful candidates to describe 
the quantum aspects of gravity.

Finally, we remark that 
according to the investigations in $F(R)$ gravity, 
when the cosmological bounce happens in the Einstein frame, 
the cosmic acceleration (inflation) may occur 
in the corresponding string frame 
via the conformal transformation~\cite{Odintsov:2014gea}. 
Thus, if there exists a kind of duality between 
the bounce phenomenon in the Einstein frame 
and inflation in the string frame, and vice versa, 
by comparing the theoretical results on the spectral index 
of the curvature perturbations and the tensor-to-scalar ratio 
in the Einstein frame with their observational values 
and using such a duality, 
we can judge whether the corresponding bounce cosmology is realistic
or not (see also Ref.~\cite{Elizalde:2014uba}).


\section*{Acknowledgments}

This work was partially supported by MINECO (Spain) project FIS2010-15640 (S.D.O.),
the grant of Russian Ministry of Education and Science, project TSPU-139 and
the grant for LRSS, project No 88.2014.2 (S.D.O. and A.N.M.),
and the JSPS Grant-in-Aid
for Young Scientists (B) \# 25800136 (K.B.).




\begin{thebibliography}{99}

\bibitem{Joyce:2014kja}
  A.~Joyce, B.~Jain, J.~Khoury and M.~Trodden,
  arXiv:1407.0059 [astro-ph.CO].

\bibitem{REV-NO-CF-CD}
%
S.~Nojiri and S.~D.~Odintsov,
Phys.\ Rept.\ {\bf 505} (2011) 59
[arXiv:1011.0544 [gr-qc]];\\
%
S.~Nojiri and S.~D.~Odintsov,
eConf C {\bf 0602061} (2006) 06
[Int.\ J.\ Geom.\ Meth.\ Mod.\ Phys.\ {\bf 4} (2007) 115]
[hep-th/0601213];\\
%
   K.~Bamba and S.~D.~Odintsov,
   arXiv:1402.7114 [hep-th];\\
%
S.~Capozziello and V.~Faraoni,
\textit{Beyond Einstein Gravity}
(Springer, Dordrecht, 2010);\\
%
S.~Capozziello and M.~De Laurentis,
Phys.\ Rept.\ {\bf 509} (2011) 167
[arXiv:1108.6266 [gr-qc]]; \\
%
  A.~de la Cruz-Dombriz and D.~Saez-Gomez,
  Entropy {\bf 14} (2012) 1717
  [arXiv:1207.2663 [gr-qc]].
%

\bibitem{Bamba:2012cp}
K.~Bamba, S.~Capozziello, S.~Nojiri and S.~D.~Odintsov,
Astrophys.\ Space Sci.\ {\bf 342} (2012) 155
[arXiv:1205.3421 [gr-qc]].

\bibitem{MBC}
%
   R.~H.~Brandenberger,
   Int.\ J.\ Mod.\ Phys.\ Conf.\ Ser.\  {\bf 01} (2011) 67
   [arXiv:0902.4731 [hep-th]];\\
%
   R.~H.~Brandenberger,
   AIP Conf.\ Proc.\  {\bf 1268} (2010) 3
   [arXiv:1003.1745 [hep-th]];\\
%
   R.~H.~Brandenberger,
   PoS ICFI {\bf 2010} (2010) 001
   [arXiv:1103.2271 [astro-ph.CO]].
%

\bibitem{Brandenberger:2012zb}
   R.~H.~Brandenberger,
   arXiv:1206.4196 [astro-ph.CO].
%

\bibitem{Novello:2008ra}
   M.~Novello and S.~E.~P.~Bergliaffa,
   Phys.\ Rept.\  {\bf 463}, 127 (2008)
   [arXiv:0802.1634 [astro-ph]].

\bibitem{Erickson:2003zm}
   J.~K.~Erickson, D.~H.~Wesley, P.~J.~Steinhardt and N.~Turok,
   Phys.\ Rev.\ D {\bf 69} (2004) 063514
   [hep-th/0312009].

\bibitem{Belinsky:1970ew}
   V.~A.~Belinsky, I.~M.~Khalatnikov and E.~M.~Lifshitz,
   Adv.\ Phys.\  {\bf 19}, 525 (1970).

\bibitem{B-P}
%
   B.~Xue and P.~J.~Steinhardt,
   Phys.\ Rev.\ Lett.\  {\bf 105} (2010) 261301
   [arXiv:1007.2875 [hep-th]];\\
%
   B.~Xue and P.~J.~Steinhardt,
   Phys.\ Rev.\ D {\bf 84} (2011) 083520
   [arXiv:1106.1416 [hep-th]];\\
%
   Y.~-F.~Cai, D.~A.~Easson and R.~Brandenberger,
   JCAP {\bf 1208} (2012) 020
   [arXiv:1206.2382 [hep-th]];\\
%
   Y.~-F.~Cai, R.~Brandenberger and P.~Peter,
   Class.\ Quant.\ Grav.\  {\bf 30} (2013) 075019
   [arXiv:1301.4703 [gr-qc]];\\
%
   T.~Qiu, X.~Gao and E.~N.~Saridakis,
   Phys.\ Rev.\ D {\bf 88} (2013) 043525
   [arXiv:1303.2372 [astro-ph.CO]].
%

\bibitem{Khoury:2001wf}
   J.~Khoury, B.~A.~Ovrut, P.~J.~Steinhardt and N.~Turok,
   Phys.\ Rev.\ D {\bf 64} (2001) 123522
   [hep-th/0103239].

\bibitem{Cai:2013kja}
   Y.~-F.~Cai, E.~McDonough, F.~Duplessis and R.~H.~Brandenberger,
   JCAP {\bf 1310} (2013) 024
   [arXiv:1305.5259 [hep-th]].

\bibitem{Prof-Piao}
%
  Z.~-G.~Liu, Z.~-K.~Guo and Y.~-S.~Piao,
  Phys.\ Rev.\ D {\bf 88}, 063539 (2013)
  [arXiv:1304.6527 [astro-ph.CO]]; \\ 
%
  Y.~T.~Wang and Y.~S.~Piao,
  Phys.\ Lett.\ B {\bf 741} (2015) 55
  [arXiv:1409.7153 [gr-qc]].
%

\bibitem{B-F(R)}
%
  C.~Barragan, G.~J.~Olmo and H.~Sanchis-Alepuz,
  Phys.\ Rev.\ D {\bf 80} (2009) 024016
  [arXiv:0907.0318 [gr-qc]];\\
%
  T.~Saidov and A.~Zhuk,
  Phys.\ Rev.\ D {\bf 81} (2010) 124002
  [arXiv:1002.4138 [hep-th]];\\
%
  C.~Barragan and G.~J.~Olmo,
  Phys.\ Rev.\ D {\bf 82} (2010) 084015
  [arXiv:1005.4136 [gr-qc]];\\
%
   M.~Bouhmadi-Lopez, J.~Morais and A.~B.~Henriques,
   Phys.\ Rev.\ D {\bf 87} (2013) 103528
   [arXiv:1210.1761 [astro-ph.CO]];\\
%
  G.~Leon and A.~A.~Roque,
  JCAP {\bf 1405} (2014) 032
  [arXiv:1308.5921 [astro-ph.CO]];\\ 
%
   G.~Leon and A.~A.~Roque,
   arXiv:1308.5921 [astro-ph.CO];\\
%
   K.~Bamba, A.~N.~Makarenko, A.~N.~Myagky, S.~Nojiri and S.~D.~Odintsov,
   JCAP {\bf 1401} (2014) 008
   [arXiv:1309.3748 [hep-th]].
%

\bibitem{Bamba:2014mya}
  K.~Bamba, A.~N.~Makarenko, A.~N.~Myagky and S.~D.~Odintsov,
  Phys.\ Lett.\ B {\bf 732} (2014) 349
  [arXiv:1403.3242 [hep-th]].

\bibitem{Cai:2011tc}
  Y.~F.~Cai, S.~H.~Chen, J.~B.~Dent, S.~Dutta and E.~N.~Saridakis,
  Class.\ Quant.\ Grav.\  {\bf 28} (2011) 215011
  [arXiv:1104.4349 [astro-ph.CO]].

\bibitem{Cai:2012ag}
  Y.~F.~Cai, C.~Gao and E.~N.~Saridakis,
  JCAP {\bf 1210} (2012) 048
  [arXiv:1207.3786 [astro-ph.CO]].

%
\bibitem{Olmo:2008nf}
   G.~J.~Olmo and P.~Singh,
   JCAP {\bf 0901} (2009) 030
   [arXiv:0806.2783 [gr-qc]].
%

\bibitem{Odintsov:2014gea}
  S.~D.~Odintsov and V.~K.~Oikonomou,
  Phys.\ Rev.\ D {\bf 90} (2014) 12,  124083
  [arXiv:1410.8183 [gr-qc]].

\bibitem{Refs-LQC}
%
  M.~Sami, P.~Singh and S.~Tsujikawa,
  Phys.\ Rev.\ D {\bf 74} (2006) 043514
  [gr-qc/0605113];\\
%
  P.~Singh, K.~Vandersloot and G.~V.~Vereshchagin,
  Phys.\ Rev.\ D {\bf 74} (2006) 043510
  [gr-qc/0606032];\\
%
  A.~Ashtekar, T.~Pawlowski and P.~Singh,
  Phys.\ Rev.\ D {\bf 74} (2006) 084003
  [gr-qc/0607039];\\
%
  A.~Ashtekar,
  Nuovo Cim.\ B {\bf 122} (2007) 135
  [gr-qc/0702030];\\
%
  D.~Samart and B.~Gumjudpai,
  Phys.\ Rev.\ D {\bf 76} (2007) 043514
  [arXiv:0704.3414 [gr-qc]];\\
%
  T.~Naskar and J.~Ward,
  Phys.\ Rev.\ D {\bf 76} (2007) 063514
  [arXiv:0704.3606 [gr-qc]];\\
%
  E.~J.~Copeland, D.~J.~Mulryne, N.~J.~Nunes and M.~Shaeri,
  Phys.\ Rev.\ D {\bf 77} (2008) 023510
  [arXiv:0708.1261 [gr-qc]];\\
%
  M.~Bojowald,
  Class.\ Quant.\ Grav.\  {\bf 26} (2009) 075020
  [arXiv:0811.4129 [gr-qc]];\\
%
  P.~Singh,
  Class.\ Quant.\ Grav.\  {\bf 26} (2009) 125005
  [arXiv:0901.2750 [gr-qc]];\\
%
  A.~Corichi and P.~Singh,
  Phys.\ Rev.\ D {\bf 80} (2009) 044024
  [arXiv:0905.4949 [gr-qc]];\\
%
  Y.~F.~Cai, R.~Brandenberger and X.~Zhang,
  JCAP {\bf 1103} (2011) 003
  [arXiv:1101.0822 [hep-th]];\\
%
  Y.~F.~Cai, S.~H.~Chen, J.~B.~Dent, S.~Dutta and E.~N.~Saridakis,
  Class.\ Quant.\ Grav.\  {\bf 28} (2011) 215011
  [arXiv:1104.4349 [astro-ph.CO]];\\
%
  Y.~F.~Cai, R.~Brandenberger and X.~Zhang,
  Phys.\ Lett.\ B {\bf 703} (2011) 25
  [arXiv:1105.4286 [hep-th]];\\
%
  A.~Ashtekar and P.~Singh,
  Class.\ Quant.\ Grav.\  {\bf 28} (2011) 213001
  [arXiv:1108.0893 [gr-qc]];\\
%
  T.~Cailleteau, J.~Mielczarek, A.~Barrau and J.~Grain,
  Class.\ Quant.\ Grav.\  {\bf 29} (2012) 095010
  [arXiv:1111.3535 [gr-qc]];\\
%
  T.~Cailleteau, A.~Barrau, J.~Grain and F.~Vidotto,
  Phys.\ Rev.\ D {\bf 86} (2012) 087301
  [arXiv:1206.6736 [gr-qc]];\\
%
  E.~Wilson-Ewing,
  JCAP {\bf 1303} (2013) 026
  [arXiv:1211.6269 [gr-qc]];\\
%
  J.~Amor\'{o}s, J.~de Haro and S.~D.~Odintsov,
  Phys.\ Rev.\ D {\bf 87} (2013) 104037
  [arXiv:1305.2344 [gr-qc]];\\
%
  Y.~F.~Cai and E.~Wilson-Ewing,
  JCAP {\bf 1403} (2014) 026
  [arXiv:1402.3009 [gr-qc]];\\
%
  J.~Amor\'{o}s, J.~de Haro and S.~D.~Odintsov,
  Phys.\ Rev.\ D {\bf 89} (2014) 104010
  [arXiv:1402.3071 [gr-qc]];\\
%
  C.~Li, R.~H.~Brandenberger and Y.~K.~E.~Cheung,
  Phys.\ Rev.\ D {\bf 90} (2014) 12,  123535
  [arXiv:1403.5625 [gr-qc]];\\
%
  J.~de Haro,
  Europhys.\ Lett.\  {\bf 107} (2014) 29001
  [arXiv:1403.4529 [gr-qc]];\\
%
  J.~de Haro and J.~Amor\'{o}s,
  JCAP {\bf 1408} (2014) 025
  [arXiv:1403.6396 [gr-qc]];\\
%
  J.~Haro and J.~Amoros,
  JCAP {\bf 1412} (2014) 12,  031
  [arXiv:1406.0369 [gr-qc]];\\ 
%
  J.~Quintin, Y.~F.~Cai and R.~H.~Brandenberger,
  Phys.\ Rev.\ D {\bf 90} (2014) 063507
  [arXiv:1406.6049 [gr-qc]].
%

\bibitem{Ade:2014xna}
  P.~A.~R.~Ade {\it et al.}  [BICEP2 Collaboration],
  Phys.\ Rev.\ Lett.\  {\bf 112} (2014) 241101
  [arXiv:1403.3985 [astro-ph.CO]].

\bibitem{Ade:2015tva}
  P.~A.~R.~Ade {\it et al.}  [BICEP2 and Planck Collaborations],
  [arXiv:1502.00612 [astro-ph.CO]].

\bibitem{Planck:2015xua}
  P.~A.~R.~Ade {\it et al.}  [Planck Collaboration],
  arXiv:1502.01589 [astro-ph.CO].

\bibitem{Ade:2015rim}
  P.~A.~R.~Ade {\it et al.}  [XXX Collaboration],
  arXiv:1502.01590 [astro-ph.CO].

\bibitem{Ade:2015lrj}
  P.~A.~R.~Ade {\it et al.}  [Planck Collaboration],
  arXiv:1502.02114 [astro-ph.CO].

\bibitem{Cai:2014xxa}
  Y.~F.~Cai, J.~Quintin, E.~N.~Saridakis and E.~Wilson-Ewing,
  JCAP {\bf 1407} (2014) 033
  [arXiv:1404.4364 [astro-ph.CO]].

\bibitem{Elizalde:2014uba}
  E.~Elizalde, J.~de Haro and S.~D.~Odintsov,
  arXiv:1411.3475 [gr-qc].

\bibitem{A-G}
%
  I.~Bars, S.~H.~Chen, P.~J.~Steinhardt and N.~Turok,
  Phys.\ Lett.\ B {\bf 715} (2012) 278
  [arXiv:1112.2470 [hep-th]]; \\
%
  I.~Bars, P.~Steinhardt and N.~Turok,
  Phys.\ Rev.\ D {\bf 89} (2014) 061302
  [arXiv:1312.0739 [hep-th]]; \\
%
  K.~Bamba, S.~Nojiri, S.~D.~Odintsov and D.~S\'{a}ez-G\'{o}mez,
  Phys.\ Lett.\ B {\bf 730} (2014) 136
  [arXiv:1401.1328 [hep-th]]; \\
%
  V.~K.~Oikonomou and N.~Karagiannakis,
  J.\ Grav.\  {\bf 2014} (2014) 625836
  [arXiv:1408.0398 [gr-qc]].
%

\bibitem{KS-KSS}
%
  S.~Kawai and J.~Soda,
  Phys.\ Lett.\ B {\bf 460} (1999) 41
  [gr-qc/9903017];\\ 
%
  S.~Kawai, M.~a.~Sakagami and J.~Soda,
  Phys.\ Lett.\ B {\bf 437} (1998) 284
  [gr-qc/9802033].
%

\bibitem{Kawai:1999xn}
  S.~Kawai and J.~Soda,
  gr-qc/9906046.

\bibitem{Antoniadis:1993jc}
  I.~Antoniadis, J.~Rizos and K.~Tamvakis,
  Nucl.\ Phys.\ B {\bf 415} (1994) 497
  [hep-th/9305025].

\bibitem{Cartier}
C.~Cartier, J.~Hwang and E.~J.~Copeland,
Phys.\ Rev.\ D {\bf 64} (2001) 103504
[astro-ph/0106197].

\bibitem{Lidsey:1999mc}
  J.~E.~Lidsey, D.~Wands and E.~J.~Copeland,
  Phys.\ Rept.\  {\bf 337} (2000) 343
  [hep-th/9909061].

\bibitem{Metsaev:1987zx}
  R.~R.~Metsaev and A.~A.~Tseytlin,
  Nucl.\ Phys.\ B {\bf 293} (1987) 385.

\bibitem{Copeland:1997ug}
  E.~J.~Copeland, R.~Easther and D.~Wands,
  Phys.\ Rev.\ D {\bf 56} (1997) 874
  [hep-th/9701082].

\bibitem{FMS-BM}
%
  S.~Foffa, M.~Maggiore and R.~Sturani,
  Nucl.\ Phys.\ B {\bf 552} (1999) 395
  [hep-th/9903008];\\ 
%
  R.~Brustein and R.~Madden,
  JHEP {\bf 9907} (1999) 006
  [hep-th/9901044].
%

\bibitem{Linde:1981mu}
  A.~D.~Linde,
  Phys.\ Lett.\ B {\bf 108} (1982) 389.

\bibitem{Albrecht:1982wi}
  A.~Albrecht and P.~J.~Steinhardt,
  Phys.\ Rev.\ Lett.\  {\bf 48} (1982) 1220.

\bibitem{Freese:1990rb}
  K.~Freese, J.~A.~Frieman and A.~V.~Olinto,
  Phys.\ Rev.\ Lett.\  {\bf 65} (1990) 3233.

\bibitem{Callan:1985ia}
  C.~G.~Callan, Jr., E.~J.~Martinec, M.~J.~Perry and D.~Friedan,
  Nucl.\ Phys.\ B {\bf 262} (1985) 593.

\bibitem{A-E-N}
%
  I.~Antoniadis, E.~Gava and K.~S.~Narain,
  Phys.\ Lett.\ B {\bf 283} (1992) 209
  [hep-th/9203071];\\ 
%
  I.~Antoniadis, E.~Gava and K.~S.~Narain,
  Nucl.\ Phys.\ B {\bf 383} (1992) 93
  [hep-th/9204030].
%

\bibitem{Rizos:1993rt}
  J.~Rizos and K.~Tamvakis,
  Phys.\ Lett.\ B {\bf 326} (1994) 57
  [gr-qc/9401023]. 

\bibitem{Easther:1996yd}
  R.~Easther and K.~I~Maeda,
  Phys.\ Rev.\ D {\bf 54} (1996) 7252
  [hep-th/9605173].

\bibitem{M-FM-F}
%
  K.~I.~Maeda,
  Phys.\ Rev.\ D {\bf 39} (1989) 3159; \\
%
Y.~Fujii and K.~I.~Maeda,
\textit{The Scalar-Tensor Theory of Gravitation}
(Cambridge University Press, Cambridge, United Kingdom, 2003); \\
%
V.~Faraoni,
\textit{Cosmology in scalar tensor gravity}
(Springer, 2004).

\bibitem{Maeda:2009}
K.~I.~Maeda, N.~Ohta, and Y.~Sasagawa,
Phys.\ Rev.\ D {\bf 80} (2009) 104032
[arXiv:0908.4151[hep-th]].


\end{thebibliography}
\end{document}